\documentclass[11pt]{amsart}
\usepackage{graphicx}

\newtheorem{thm}{Theorem}[section]

\newtheorem{lem}[thm]{Lemma}
\newtheorem{prop}[thm]{Proposition}

\theoremstyle{definition}
\newtheorem{defn}[thm]{Definition}
\theoremstyle{remark}
\newtheorem{rem}[thm]{Remark}
\newtheorem{ex}[thm]{Example}
\numberwithin{equation}{section}

\newcommand{\abs}[1]{\left\vert#1\right\vert}

\newcommand{\Real}{\mathbb R}
\newcommand{\Natural}{\mathbb N}

\newcommand{\A}{\mathcal{A}}
\newcommand{\B}{\mathcal{B}}

\newcommand{\such}{\kern0.17em | \kern0.17em}
\newcommand{\ud}{\mathrm d}

\newcommand{\prob}{\mathbb{P}}
\newcommand{\qprob}{\mathbb{Q}}
\newcommand{\expec}{\mathbb{E}}

\newcommand{\filtration}{\mathbf{F} = \pare{\mathcal{F}_t}_{t \in \Real_+}}
\newcommand{\F}{\mathcal{F}}

\newcommand{\cadlag}{c\`adl\`ag }

\newcommand{\p}{\mathrm{p}}

\newcommand{\g}{\mathfrak{g}}
\newcommand{\rel}{\mathfrak{rel}}

\newcommand{\X}{\Pi}

\newcommand{\pare}[1]{\left(#1\right)}
\newcommand{\bra}[1]{\left[#1\right]}
\newcommand{\dbra}[1]{[\kern-0.15em[ #1 ]\kern-0.15em]}
\newcommand{\N}{\mathfrak{N}}

\newcommand{\C}{\mathfrak{C}}

\newcommand{\St}{\widetilde{S}}

\newcommand{\K}{\mathfrak{K}}

\newcommand{\Cc}{\check{\mathfrak{C}}}

\newcommand{\I}{\mathfrak{I}}

\newcommand{\Exp}{\mathcal E}
\newcommand{\indic}{\mathbb{I}}

\newcommand{\supp}{\mathsf{supp}}

\newcommand{\Def}{\mathfrak{D}}
\newcommand{\cone}{\mathsf{cone}}

\newcommand{\num}{num\'eraire }
\newcommand{\hx}{x \indic_{\{| x | \leq 1 \}}}
\newcommand{\hbarx}{x \indic_{\{| x | > 1 \}}}

\newcommand{\proj}{\mathsf{pr}}

\newcommand{\EMM}{\emph{EMM }}

\newcommand{\NUIPC}{NUIP$_\C$ }
\newcommand{\NFLVRC}{NFLVR$_\C$ }
\newcommand{\NUPBRC}{NUPBR$_\C$ }
\newcommand{\NAC}{NA$_\C$ }
\newcommand{\ESMMC}{ESMM$_\C$ }
\newcommand{\ESMDC}{ESMD$_\C$ }

\newcommand{\QprobC}{\mathfrak{Q}_\C}

\begin{document}

\title[No-Free-Lunch equivalences for Exponential L\'evy models]{No-Free-Lunch equivalences for Exponential L\'evy models under Convex Constraints on Investment}%
\author[Constantinos Kardaras]{Constantinos Kardaras}
\address{Mathematics and Statistics Department, Boston University, MA02215}
\email{kardaras@bu.edu}

\thanks{The author would like to thank two anonymous referees for their valuable input.}%

\date{\today}%

\begin{abstract}
We provide equivalence of numerous no-free-lunch type conditions for
financial markets where the asset prices are modeled as exponential
L\'evy processes, under possible convex constraints in the use of
investment strategies. The general message is the following: if any
kind of free lunch exists in these models it has to be of the most
egregious type, generating an \emph{increasing} wealth. Furthermore,
we connect the previous to the existence of the \textsl{\num
portfolio}, both for its particular expositional clarity in
exponential L\'evy models and as a first step in obtaining analogues
of the no-free-lunch equivalences in general semimartingale models,
a task that is taken on in Karatzas and Kardaras \cite{KK: num and
arbitrage}.
\end{abstract}

\maketitle


\setcounter{section}{-1}

\section{Introduction} \label{sec: Introduction and some Notation}

\subsection{Discussion}

An \textsl{exponential L\'evy process} --- as its name suggests
--- is simply the exponential 
of a L\'evy process. Models of financial markets that assume an
exponential L\'evy structure for the movement of the stock-price
processes have become increasingly popular in the last years, partly
because of their analytical tractability (since their distributional
properties are uniquely determined by their L\'evy triplet) and
partly because they provide a reasonably good fit to actual
financial data. Noteworthy examples are the four-parameter
\textsl{CGMY} model of Carr, Geman, Madan and Yor \cite{CGMY} and
the \textsl{hyperbolic model} of Eberlein, Keller and Prause
\cite{Ebe-Ke-Pra}. One effect of this popularity is the
proliferation of academic courses that include in their teaching
curriculum models of this sort.

It is somewhat of \emph{folklore} that if free lunches exist in
exponential L\'evy models, they are of the most egregious form: one
can invest in a way so to obtain an \emph{increasing} wealth
process. As a result, many subtle differences existing in different
formulations of a ``free lunch'' definition in more general models
disappear, something with both good and bad consequences. On the
positive side, one can provide a proof of the Fundamental Theorem of
Asset Pricing (FTAP) with minimal effort that can be easily taught
--- in particular, no functional-analytic background is required and
the proof uses reasonably standard facts from L\'evy-process theory.
The offset is that mere knowledge of the no-free-lunch situation in
exponential L\'evy models is inadequate to provide the whole picture
and complications that prevail in semimartingale models.

The purpose of this paper is twofold: first, to provide a quick and
easy proof\footnote{In view of remarks and questions that arose
during presentations of the material stemming from the author's Ph.D
thesis \cite{K: thesis}, it became clear that there was a desire for
a self-contained treatment of the FTAP for exponential L\'evy
models. In this respect, one of the main results (Theorem \ref{thm:
No Arbitrage for Levy finite time constrained}) is \emph{dedicated}
to those who expressed interest for it, with the hope that it will
help their teaching.} of the above folklore fact for multi-asset,
finite-time horizon models under convex constraints, establishing
many equivalences regarding no-free-lunch notions and
(super)martingale measures; second, to explore the structure of the
so-called \textsl{\num portfolio} in the context of exponential
L\'evy models to the extend where the transition to general
semimartingale models will be possible.

\smallskip

Economic agents typically face restrictions in the free use of
portfolios --- a first example being short-sale constraints. In this
``constrained'' setting and in the context of the FTAP, one cannot
claim any more that no-free-lunch criteria expressed in terms of the
restricted collection of admissible strategies imply existence of
equivalent martingale measures for the stock-price process. An
extreme example is total prevention for the use of any portfolio,
except keeping all the wealth in the savings account; in this
trivial case even if free lunches exist in the \emph{unconstrained}
market, they cannot be used because the agent \emph{cannot} invest
in them. To compensate for the fact that we are \emph{only}
considering constrained strategies, we have to introduce notions of
equivalent probability measures that act only on the wealth-process
class and not on the stock-price process. As long as the constraints
are of the form of a convex \emph{cone}, the concept of
\textsl{equivalent supermartingale measure} (ESMM --- also called
\textsl{separating measure} in the literature) does the trick: we
have to make sure that under an equivalent change of probability all
wealth processes are supermartingales.

The first main result of the paper --- a version of the FTAP for
convex-cone-constrained exponential L\'evy models --- is Theorem
\ref{thm: No Arbitrage for Levy finite time constrained}. The
``difficult'' part of the proof of Theorem \ref{thm: No Arbitrage
for Levy finite time constrained} follows the idea of Rogers
\cite{Rogers}: solve a utility maximization problem and construct an
ESMM using the marginal utility evaluated on the optimal wealth
process as density. Rogers implemented this for the discrete-time
case; an inductive construction based on the simple static
one-time-period model had to be utilized in order to fully prove the
FTAP for multi time-period models. Unfortunately, this construction
does not carry to general continuous-time models, exactly because
this inductive step cannot be carried over. Nevertheless, one can
use this idea when L\'evy processes are involved, because of their
``independent and stationary increments'' structure. Theorem
\ref{thm: No Arbitrage for Levy finite time constrained} presents
seven equivalences involving equivalent supermartingale measures
that respect the exponential L\'evy structure, several no-free-lunch
notions and --- most importantly
--- a condition that involves only the characteristic triplet of the
generating L\'evy process. Let us note that simpler statements than
that of Theorem \ref{thm: No Arbitrage for Levy finite time
constrained}, dealing only with equivalences for the one-stock,
unconstrained case, have already appeared in Jacub\.enas
\cite{Jacub: option pricing}, Cherny and Shiryaev \cite{Cherny :
Levy process}, as well as in Selivanov \cite{Selivanov}. The proof
contained in the first paper is inspired by the work of Eberlein and
Jacod \cite{Ebe-Jac: range of option prices} and the proof in the
other papers more or less use the idea of the Esscher transform (as
we do here); the proofs sometimes are slightly more complicated and
--- as already mentioned --- are valid in a one-dimensional, unconstrained setting.

\smallskip

It seems reasonable to proceed in proving analogous no-free-lunch
equivalences in the case we have convex, but \emph{not necessarily
conic} constraints. The moment that we try to do so, we face an
unexpected barrier: \emph{no-free-lunch conditions are no longer
sufficient to provide us with an equivalent supermartingale measure}
--- this slightly surprising fact in illustrated in Example \ref{ex:
NA does not imply ESMM}. In the quest for finding an appropriate
version of the FTAP under convex constraints we have to depart from
the world of equivalent supermartingale measures and enter the realm
of \textsl{equivalent supermartingale deflators}, i.e.,
state-price-density processes that are only supermartingales (and
not martingales) and can therefore \emph{lose mass}. A particularly
efficient way to obtain an equivalent supermartingale deflator is by
use of the \num portfolio: this is a special strategy that generates
a wealth process in such a way that relative wealth processes
generated by all other portfolios with respect to it are
supermartingales. When the \num portfolio exists, so do equivalent
supermartingale deflators --- the interesting fact is that the
converse also holds: existence of at least one equivalent
supermartingale deflator will imply that the \num portfolio exists.
This also turns out to be equivalent to requiring that the terminal
values of all wealth processes that start from unit capital are
bounded in probability, and it is exactly that last no-free-lunch
notion (which we baptize \textsl{no unbounded profit with bounded
risk}) that is tailor-made for the case of convex constraints in
order to obtain an equivalent of the FTAP. We state this result as
Theorem \ref{thm: num iff def-non-empty iff NUPBR iff NUIP}; its
proof is more technical than that of Theorem \ref{thm: No Arbitrage
for Levy finite time constrained} and its backbone is Lemma
\ref{lem: Levy charact of num}, whose proof is the whole the purpose
of section \ref{sec: crucial lemma}. To the best of the author's
knowledge, \emph{no} result of this type (dealing with convex but
not necessarily conic constraints) has appeared before in the
literature. We note that the decision to single out the statement
and proof of Lemma \ref{lem: Levy charact of num} is made not only
for presentation reasons; it will also be used in a crucial way in
Karatzas and Kardaras \cite{KK: num and arbitrage}, where a study of
the general semimartingale case is made. We further note that a
solution to the problem of maximizing expected log-utility (which is
very closely related to the \num portfolio, as is also discussed in
subsection \ref{subsec: log-optimality}) in a general semimartingale
model and under convex constraints has been carried by Goll and
Kallsen \cite{Goll - Kallsen: log-optimal}.

\smallskip

Let us mention two more results that appear in the text. First,
\textsl{completeness} for multi-asset exponential L\'evy models is
considered in subsection \ref{subsec: completeness} --- because it
is not the main point of this paper, the treatment is very brief.
Second, a result concerning the infinite-time horizon case is given
--- Theorem \ref{thm: No Arbitrage for inf-horizon Levy}. If
existence of free lunches is the \emph{exception} when dealing with
finite-time planning horizon, since it happens in the most severe
way, it is the \emph{rule} in infinite-time horizon models: one is
\emph{always} able to construct a free lunch, provided that the
\emph{original} probability is not a supermartingale measure. In the
one-dimensional case, a statement of this last result appears and is
proved in Selivanov \cite{Selivanov}; nevertheless, it is not clear
how to transfer the statement appearing there to the
multi-dimensional case that we are dealing here.

\subsection{Organization of the paper} This Introduction continues with
fixing notation and discussing basic facts concerning L\'evy
processes. Section \ref{sec: The Market, Investments and
Constraints} introduces the financial market with exponential L\'evy
discounted stock-price processes and describes wealth processes as
well as constraints. In section \ref{sec: nfl equiv, unconstrained
and cone-constrained} we introduce the no-free-lunch and
equivalent-(super)martingale notions that shall be used in the
sequel and we present the first main result: Theorem \ref{thm: No
Arbitrage for Levy finite time constrained}, that provides
equivalences for the cone-constrained case. We proceed in section
\ref{sec: Numeraire portfolio} to introduce the \num portfolio,
equivalent supermartingale deflators; then, we state Theorem
\ref{thm: num iff def-non-empty iff NUPBR iff NUIP} that covers
no-free-lunch equivalences for cone-constrained models. In the same
section we provide a result concerning the infinite-time horizon
case (Theorem \ref{thm: No Arbitrage for inf-horizon Levy}) and
discuss the connection between the \num portfolio and the
growth-optimal, that actually give us the way to \emph{construct}
it. Section \ref{sec: crucial lemma} contains only the statement and
proof of Lemma \ref{lem: Levy charact of num} that is needed to
complete the proof of Theorem \ref{thm: num iff def-non-empty iff
NUPBR iff NUIP}. We also include an Appendix with some special
results on L\'evy processes that are needed in the main text.

\subsection{Some notation}

The transpose of a vector $x \in \Real^d$ is denoted by $x^\top$,
its \textsl{norm} is $| x | := \sqrt{x^\top x}$, and superscripts
denote coordinates: $x = (x^1, \ldots x^d)^\top$. The
\textsl{indicator function} of a set $A$ is denoted by $\indic_A$;
for subsets of $\Real^d$, we write ${\{ | x | > 1 \}}$ to actually
express $\{ x \in \Real^d \such | x | > 1 \}$.

We are working on a \textsl{stochastic basis} $(\Omega, \F,
\mathbf{F}, \prob)$, where the \textsl{filtration} $\filtration$ is
right-continuous and augmented by all $\prob$-null sets. The symbol
$\expec$ always denotes \textsl{expectation} of random variables
under $\prob$. Expectations with respect to other probability
measures (say, $\qprob$) will involve the measure appearing as
superscript on $\expec$ (say, $\expec^\qprob$).

For a $d$-dimensional semimartingale $X$ and a $d$-dimensional
predictable\footnote{The \textsl{predictable $\sigma$-algebra} is
generated by all the adapted, left-continuous processes.} process
$\pi$, we shall denote by $\pi \cdot X$ the \emph{vector} stochastic
integral process whenever this makes sense, i.e., when $\pi$ is
\textsl{$X$-integrable}. One can check for example Jacod and
Shiryaev \cite{Jacod - Shiryaev} for these notions.

Any \cadlag (adapted, right-continuous with left-hand limits)
process $Z$ has an obviously-defined left-continuous --- thus
predictable
--- version $Z_-$; for concreteness, we set $Z_-(0) = 0$. We also
define the \textsl{jump process} $\Delta Z := Z - Z_-$.

Finally, for a one-dimensional semimartingale $Y$, $\Exp(Y)$ will
denote the \textsl{stochastic exponential} of $Y$, i.e.,
\emph{unique} semimartingale  $Z$ that solves the stochastic
differential equation $\ud Z_t = Z_{t-} \ud Y_t$.

\subsection{Basics of L\'evy processes}

The are several good books that one can obtain information on L\'evy
processes --- for example, Sato \cite{Sato} is a good reference for
the theoretical part, while Cont and Tankov \cite{Cont-Tankov: Levy}
provide applications in financial modeling.

Given a stochastic basis $(\Omega, \F, \mathbf{F}, \prob)$, a
$d$-dimensional \cadlag process $L$ with $L_0 = 0$, such that for
all $0 \leq s < t$, the increment $L_t - L_s$ is independent of
$\F_s$ and its distribution only depends on $t-s$ will be called an
$\mathbf{F}$-\textsl{L\'evy process}.

With a L\'evy process $L$ comes its \textsl{L\'evy triplet} $(b_L,
c_L, \nu_L)$. Here, $b_L \in \Real^d$, $c_L$ is a
nonnegative-definite $d \times d$ matrix (if $d=1$ this just reads
$c_L \in \Real_+$), and $\nu_L$ is a \textsl{L\'evy measure} on
$\Real^d$ with its Borel $\sigma$-algebra, i.e., $\nu_L$ satisfies
$\nu_L (\{0\}) = 0$ and $\int_{\Real^d} (1 \wedge |x|^2) \nu_L (\ud
x) < + \infty$ (the wedge ``$\wedge$'' denotes minimum: $f \wedge g
= \min \{f,g \}$). The finite-dimensional distributions of $L$ are
completely determined by its L\'evy triplet via the characteristic
functions
\begin{equation} \label{eq: char fun of Levy}
\expec  \exp \Big( i \sum_{j=1}^n u^\top_j (L_{t_j} - L_{t_{j-1}}) \Big) = \prod_{j=1}^n \exp
\big( (t_j - t_{j-1}) \phi (u_j) \big),
\end{equation}
for all $0 = t_0 < \ldots < t_n$ and $u_j \in \Real^d$ for all $j =1, \ldots, n$, where $i =
\sqrt{-1}$ and
\begin{equation} \label{eq: cummulant of Levy}
\phi (u) := i u^\top b_L - \frac{u^\top c_L u }{2} + \int_{\Real^d}
(e^{i u^\top x} - 1 - i u^\top x \indic_{ \{ | x | \leq 1\}} ) \nu_L
(\ud x)
\end{equation}
We have $\expec |L_t| < \infty$ for all $t
\in \Real_+$ if and only if $\int_{\Real^d} | x | \indic_{ \{ | x |
> 1\}} \nu_L (\ud x) < \infty$; then
\begin{equation} \label{eq: lin formula}
\expec L_t = t \Big(b_L + \int_{\Real^d} x \indic_{ \{ | x | > 1\}}
\nu_L (\ud x) \Big).
\end{equation}
In the one-dimensional case $d=1$, formally setting $u = - i$ in (\ref{eq: char fun of Levy}) and
(\ref{eq: cummulant of Levy}) one obtains the \emph{exponential formula} (written in logarithmic
form to ease reading):
\begin{equation} \label{eq: expo formula}
\log \big( \expec e^{L_t} \big) = t \Big( b_L + \frac{c_L}{2} +
\int_{\Real} (e^x - 1 - x \indic_{ \{ | x | \leq 1\}}) \nu_L (\ud x)
\Big);
\end{equation}
this always holds, in the sense that one side is finite if and only if the other is, and when they
are finite they give the same value.

Further results on L\'evy processes that will be useful later are
collected in the Appendix.

\section{Exponential L\'evy Models of Financial Markets} \label{sec: The Market, Investments and Constraints}

\subsection{The financial market model}

The prices of $d$ financial assets are modeled as $d$ strictly
positive semimartingales $\St^1, \ldots, \St^d$. There is also
another process $\St^0$ which models the \textsl{money market} and
plays the role of a ``benchmark'', in the sense that wealth
processes will be quoted in units of $\St^0$. We then define the
\textsl{discounted price processes} $S^i := \St^i / \St^0$ for $i=0,
\ldots, d$. The $d$-dimensional vector process $(S^1, \ldots, S^d)$
will be denoted by $S$.

We now enforce more structure on each of the discounted
price-processes; in particular, we assume that they satisfy $\ud
S^i_t = S^i_{t-} \ud X^i_t$, or equivalently $S^i = S^i_0
\Exp(X^i)$, where for all $i = 1, \ldots, d$, $X^i$ is a L\'evy
process with $\Delta X^i > -1$ (remember that $\Exp$ is the
stochastic exponential operator). Denote by $X$ the $d$-dimensional
L\'evy process $(X^1, \ldots, X^d)$. According to the
\emph{L\'evy-It\^o path decomposition} one can write
\begin{equation} \label{eq: canonical representation}
X_t = b t + \sigma \beta_t + \int_0^t \int_{\Real^d} \hx \big( \mu(\ud x, \ud u) - \nu (\ud x) \ud
u \big) + \int_0^t \int_{\Real^d} \hbarx \mu(\ud x, \ud u).
\end{equation}
With $c := \sigma \sigma^\top$, $(b,c,\nu)$ is the L\'evy triplet of
$X$. Here, $\beta$ is a standard $d$-dimensional Brownian motion,
and $\mu$ is the \textsl{jump measure} of $X$, i.e., the random
counting measure defined for $t \in \Real_+$ and $A \subseteq
\Real^d \setminus \{0\}$ by $\mu ([0,t] \times A) := \sum_{0 \leq s
\leq t} \indic_A (\Delta X_s)$.

Since $S^i = S^i_0 \Exp(X^i)$, one can actually write $L^i := \log
S^i$ in terms of $X^i$ as follows: $L_t^i = L^i_0 + X_t^i - c^{ii} t
/ 2 - \int_0^t \int_{\Real^d} [x - \log(1+x)] \mu(\ud x, \ud u)$; we
then observe that $L^i$ is a L\'evy process, and this is the reason
why models like the ones we are considering are called
\textsl{exponential L\'evy models}. Both the usual and the
stochastic logarithm of the asset prices are L\'evy processes; we
choose to state everything in terms of the stochastic
--- as opposed to the usual --- logarithm since it will be much more
convenient in the sequel.

\smallskip

We shall be mostly working on a finite-time horizon; only one result
(Theorem \ref{thm: No Arbitrage for inf-horizon Levy}) will be
stated for the infinite-time horizon case. We then fix a number $T
\in \Real_+$ (the \textsl{maturity}) and we denote $\dbra{0, T} :=
\Omega \times [0, T]$.

\subsection{Portfolios, wealth processes and constraints}

A financial agent starts with some strictly positive initial capital
which we normalize to be unit throughout, and can invest in the
assets by choosing a predictable, $d$-dimensional and $X$-integrable
process process $\pi$, which we shall refer to as
\textsl{portfolio}. We interpret $\pi^i_t$ as the \emph{proportion
of current wealth} invested in stock $i$ at time $t$; the remaining
proportion of wealth $\pi_0 := 1 - \sum_{i=1}^d \pi^i$ is then
invested in the money market. The wealth generated by this portfolio
is constrained to remain strictly positive at all times; going on
the red is not allowed in our model.

If $W^\pi$ denotes the \emph{discounted} wealth process obtained
following $\pi$, then $W^\pi > 0$ and thus $\Delta W^\pi_t > -
W^\pi_{t-}$. The previous interpretation for $\pi$ implies that
\begin{equation} \label{eq: wealth dynamics}
\frac{\ud W^\pi_t}{W^\pi_{t-}} \ = \ \sum_{i = 0}^d \pi^i_t
\frac{\ud S^i_t}{ S^i_{t-}} \ = \ \sum_{i =1}^d \pi^i_t \ud X^i_t \
\equiv \ \pi^\top_t \ud X_t,
\end{equation}
the second equality simply holding because $\ud S_t^0 = 0$ and $\ud
S_t^i = S^i_{t-} \ud X_t^i$.

The financial agent might be constrained further in the use of any
desired portfolio position; we model this by introducing a
\emph{closed} and \emph{convex} set $\C \subseteq \Real^d$ and
requiring that $\pi(\omega, t) \in \C$ for all $(\omega, t) \in
\dbra{0, T}$. For example, if the agent is prevented from selling
stock short, we have $\C = (\Real_+)^d$. If we further prevent
borrowing from the bank then we must also have $\pi^0 \geq 0$; in
other words we must use $\C = \{ \p \in \Real^d \such \p^i \geq 0
\text{ and } \sum_{i=1}^d \p^i \leq 1 \}$.

The constrains set $\C$ should be such that we at least give freedom
\emph{not} to invest in the stock market if the agent chooses to do
so. This should be modeled by requiring $0 \in \C$, but there might
also be \emph{degeneracy} in the market, i.e., linear dependence of
the returns of the stocks. The effect of this is that different
portfolios will produce the same wealth. To understand how this
notion should be formalized, consider two portfolios $\pi_1$ and
$\pi_2$ with $W^{\pi_1} = W^{\pi_2}$. Uniqueness of the stochastic
exponential implies $\pi_1 \cdot X = \pi_2 \cdot X$, or that $\zeta
:= \pi_2 - \pi_1$ will satisfy $\zeta \cdot X \equiv 0$, which is
easily seen to be equivalent to $\zeta \cdot \beta = 0$, $\zeta^\top
\Delta X = 0$ and $\zeta^\top b = 0$.

\begin{defn} \label{dfn: null investments}
For a L\'evy  triplet $(b,c,\nu)$, the linear subspace of
\textsl{null investments} $\N$ is defined as the set of vectors $\N
:= \big\{ \zeta \in \Real^d \such \zeta^\top c = 0,\ \nu[ \zeta^\top
x \neq 0 ] = 0  \text{ and } \zeta^\top b = 0 \big\}$.
\end{defn}

Finally, here comes the formal definition of our portfolio strategies.

\begin{defn} \label{dfn: K-constrained strategies}
Consider a convex and closed $\C \subseteq \Real^d$ such that $\N
\subseteq \C$. The class $\Pi_\C$ of all \textsl{$\C$-constrained
portfolios} is defined to consist of all predictable and
$X$-integrable processes $\pi$ such that $\pi^\top \Delta X > -1$
and $\pi(\omega, t) \in \C$ for all $(\omega, t) \in \dbra{0, T}$.
\end{defn}

\begin{rem} \label{rem: on natural constraints}
(\textsc{On Natural Constraints}) Observe that the positivity
requirement for $W^\pi$ implies $\pi^\top \Delta X \geq -1$; in
terms of the L\'evy measure $\nu$ this is equivalent to $\nu
[\pi^\top x < - 1] = 0$. In other words, the set $\C_0 := \{\p \in
\Real^d \such \nu [\p^\top x < - 1] = 0 \}$ present some
\emph{already} model-enforced constraints, \emph{regardless} of any
other constraints $\C$ enforced to agents. Thus, insofar as $\C_0
\subseteq \C$, we are basically regarding this as a unconstrained
case.

Even though we could in principle enrich the given constraints $\C$
to include the natural ones by considering $\C \cap \C_0$ we shall
\emph{not} do so --- we regard $\C$ as ``outside'' constraints.
\end{rem}

From the wealth dynamics (\ref{eq: wealth dynamics}) it follows that
for all $\pi \in \Pi_\C$ we have $W^\pi = \Exp(\pi \cdot X)$.
Observe that any \emph{constant} vector $\pi \in \C$ with
$\nu[\pi^\top x \leq -1] = 0$ can be considered as an element of
$\Pi_\C$ and that the wealth it generates is again an exponential
L\'evy process, because $\pi \cdot X = \pi^\top X$ is a L\'evy
process.

\begin{rem} \label{rem: does not hurt to consider non-degeneracy}
The assumption $\N \subseteq \C$ on the constraint set implies that
$\C = \C + \N$: indeed, for any $\pi \in \C$ and $\zeta \in \N
\subseteq \C$ we have that $n \zeta \in \C$ for any $n \in
\Natural$, thus the convex combination $(1-n^{-1}) \pi + \zeta$
belongs to $\C$ as well; since $\C$ is closed, $\pi + \zeta \in \C$.
Now, $\C$ is closed and $\N$ is a linear subspace; this means that
$\proj_{\N^\bot} \C = \C \cap \N^\bot$ is also closed in the
subspace $\N^\bot$, where $\proj_{\N^\bot}$ is the usual Euclidean
projection on $\N^\bot$, the orthogonal complement of $\N$. We
conclude that we can restrict attention to the set $\C \cap \N^\bot$
for the portfolios --- any degeneracy originally present in the
market disappears there.
\end{rem}

\section{No-Free-Lunch Equivalences for Convex-Cone-Constrained
Models} \label{sec: nfl equiv, unconstrained and cone-constrained}

\subsection{Classical free-lunch-type notions} \label{subsec: reminder of arbitrage}

We remind ourselves of some ``no free lunch'' conditions that will
be matter of our study later on.

\begin{defn} \label{dfn: arbitrage notions}
For the following three definitions we consider our financial model
with $\C$-constrained portfolio class $\Pi_\C$.

\smallskip

\noindent (1) A portfolio $\pi \in \Pi_\C$ \emph{generates an
arbitrage}, if $\prob[W^\pi_T \geq 1] = 1$ and $\prob[W^\pi_T > 1] >
0$. If no such portfolio exists we say that the $\C$-constrained
market satisfies \textsl{no arbitrage} (NA$_\C$).

\smallskip

\noindent (2) The $\C$-constrained market is said to satisfy the
\textsl{no unbounded profit with bounded risk} (NUPBR$_\C$)
condition if the collection of positive random variables
$(W_T^{\pi})_{\pi \in \Pi_\C}$ is bounded in probability, i.e., if
$\lim_{m \to \infty} \downarrow (\sup_{\pi \in \Pi_\C}
\prob[W^{\pi}_T > m]) = 0$.

\smallskip

\noindent (3) A \textsl{free lunch with vanishing risk} is a
sequence of portfolios $(\pi_n)_{n \in \Natural}$ with
$\prob[W_T^{\pi_n} \geq 1 - \delta_n] = 1$ for a decreasing sequence
$\delta_n \downarrow 0$, such that there exists $\epsilon > 0$ with
$\prob[W^{\pi_n}_T > 1 + \epsilon] > \epsilon$. If such a situation
is impossible by use of $\C$-constrained portfolios, we say that the
\textsl{no free lunch with vanishing risk} (NFLVR$_\C$) condition
holds.

\smallskip

In the unconstrained case we skip the subscripts ``$\Real^d$'' and
write NA, NUPBR and NFLVR.

\end{defn}

\NAC is the most classical of all three notions and its
interpretation is straightforward. The \NUPBRC condition says that
the probability of making ``crazy'' amounts of money at time $T$
starting from unit capital and staying positive can be estimated
uniformly over all portfolios and converges to zero as that
``crazy'' amount goes to infinity. \NFLVRC was introduced by Delbaen
and Schachermayer \cite{DS: FTAP locally bdd} in order to prove a
general version of the Fundamental Theorem of Asset Pricing. It can
be further shown that if a free lunch with vanishing risk exists, we
can choose $(W^{\pi_n}_T)_{n \in \Natural}$ so that it converges
$\prob$-a.s. to a $[1, + \infty]$-valued random variable $f$ which
will (necessarily) satisfy $\prob[f > 1] > 0$ --- then, $f$ is the
free lunch and $\delta_n$ is the downside risk of using the
portfolio $\pi_n$ which \emph{vanishes} to zero.

It is an easy exercise that \NFLVRC implies both \NAC and \NUPBRC
and we shall use this fact later on --- actually, \NFLVRC
$\Leftrightarrow$ \NAC $+$ NUPBR$_\C$ if $\C$ is a \emph{cone} (see
Karatzas and Kardaras \cite{KK: num and arbitrage}). In general
semimartingale models, none of the two conditions \NAC and \NUPBRC
implies the other, and they are not mutually exclusive; for
exponential L\'evy markets and cone constraints we shall see that
they are equivalent.

\subsection{Unbounded Increasing Profit} \label{subsec: unbounded increasing profit}
We now introduce yet another form of arbitrage --- actually, the
most egregious one: existence of wealth processes that start with
unit capital, manage to make something, and are furthermore
increasing.

\begin{defn} \label{dfn: NUIP}
Let $\check{\C} := \bigcap_{a > 0} a \C$ be the \textsl{recession
cone} of $\C$. A $\pi \in \Pi_{\Cc}$ is said to generate an
\textsl{unbounded increasing profit} if $W^\pi$ is increasing, i.e.,
if $\prob[W^\pi_s \leq W^\pi_t, \forall 0 \leq s < t \leq T] = 1$,
and if $\prob[W^\pi_T
> 1] > 0$. If no such portfolio exists we say that the \textsl{no
unbounded increasing profit} (NUIP$_\C$) condition holds.
\end{defn}

The process $W^\pi$ is increasing if and only if $\pi \cdot X$ is
increasing. The qualifier ``unbounded'' stems from the fact that
since $\pi \in \Pi_{\Cc}$, one can invest as much as one wishes on
the strategy $\pi$; by doing so, the agent's wealth will be
multiplied, and as the position becomes arbitrarily large, the gains
are unbounded.

The \NUIPC condition is the weakest ``no free lunch'' notion of them
all defined; both \NAC and \NUPBRC obviously imply it. Amazingly (or
not so amazingly --- see Lemma \ref{lem: pos is incr for Levy}) it
turns out that in exponential L\'evy markets and under cone
constraints \NUIPC is equivalent to all previously-defined arbitrage
notions. In other words, if any opportunities for free lunches exist
in exponential L\'evy models, they are of the most egregious type:
unbounded increasing profits. Of course, the reason for this is the
very special structure of exponential L\'evy models that makes many
``optimal'' portfolios (for example, the ones that correspond to
 power utility functions) constant; this has been observed and known since
the work of Foldes \cite{Foldes}.

\subsection{Immediate arbitrage opportunities}

To obtain the connection of arbitrage --- and especially the \NUIPC
condition --- with the L\'evy triplet of $X$, we now give the
definition of the immediate arbitrage opportunity vectors.

\begin{defn} \label{dfn: instant_arb_opport}
Let $(b,c,\nu)$ be any L\'evy triplet. Define the set $\I$ of
\textsl{immediate arbitrage opportunities} to be the set of vectors
$\xi \in \Real^d \setminus \N$ such that the following three
conditions hold: (1) $\xi^\top c = 0$, (2) $\nu [ \xi^\top x < 0 ] =
0$, and (3) $\xi^\top b - \int  \xi^\top \hx \nu (\ud x) \geq 0$.
\end{defn}

Observe that we are \emph{not} considering null investments in the
previous definition --- a $\xi \in \N$ satisfies the three
conditions, but cannot be considered an ``arbitrage opportunity''
since it has zero returns. It is easy to see that $\I$ is a cone
with the whole ``face'' $\N$ removed.

As Lemma \ref{lem: imm arb opp implies UIP} below will show,
immediate arbitrage opportunities are constant portfolios that
result in increasing profits. It is instructive to give examples in
two special cases of L\'evy processes, in order to also make
comparison with previous work.

\begin{ex} \label{ex: special cases of imed arb opport}

We first consider the multi-dimensional
Samuelson-Black-Scholes-Merton model, i.e., $X_t = b t + \sigma
\beta_t$. Since $\nu \equiv 0$, an immediate arbitrage opportunity
is a $\xi \in \Real^d$ with $\xi^\top c = 0$ and $\xi^\top b
> 0$. It then follows that absence of immediate arbitrage
opportunities is equivalent to the existence of $\rho \in \Real^d$
such that $b = c \rho$. The vector $\rho$ always exists if $c$ is
nonsingular.

Consider now a general one-stock exponential L\'evy model, which we
assume to be nontrivial (in that $X \neq 0$; here this is equivalent
to $\N = \{ 0 \}$). When do immediate arbitrage opportunities exist?
Observe that if there exists a diffusion component, i.e., if $c >
0$, then $\I = \emptyset$ because (1) of Definition \ref{dfn:
instant_arb_opport} fails for all $\xi \neq 0$. If $c = 0$, then we
only need to check (2) and (3) of Definition \ref{dfn:
instant_arb_opport} for $\xi = 1$ and $\xi = -1$. Now, $\xi = 1$ is
an immediate arbitrage opportunity if $\nu [x < 0] = 0$ and $b -
\int \hx \nu (\ud x) \geq 0$, and it is easy to see
--- or consult Lemma \ref{lem: imm arb opp implies UIP} to convince
yourselves --- that this is the case if and only if $X$
(equivalently, the stock price $S$) is increasing. Similarly, $\xi =
-1$ is an immediate arbitrage opportunity if and only if $X$, and
equivalently $S$, is decreasing. We thus get exactly the condition
that appears in \cite{Cherny : Levy process} and \cite{Jacub: option
pricing}.
\end{ex}

The following lemma explains the relevance of the above Definition
\ref{dfn: instant_arb_opport} with arbitrage.

\begin{lem} \label{lem: imm arb opp implies UIP}
Suppose that $\I \neq \emptyset$. Then, $\xi \in \I$ if and only if
$W^\xi$ is an increasing process and $\prob[W^\xi_T > 1] > 0$. Thus,
if further $\xi \in \check{\C}$, then $\xi$ is an unbounded
increasing profit.
\end{lem}

\proof Suppose that $\I \neq \emptyset$ and pick $\xi \in \I$.
Condition (1) of Definition \ref{dfn: instant_arb_opport} implies
that $\xi^\top \beta \equiv 0$ and condition (2) that $\pi^\top
\Delta X \geq 0$; in particular, $\pi^\top X$ will then be a L\'evy
process of finite variation and we can write
\begin{equation} \label{eq: decomp of UIP}
\xi^\top X_t = t \Big( \xi^\top b - \int_{\Real^d} \xi^\top \hx \nu(\ud x) \Big) + \int_0^t
\int_{\Real^d} (\xi^\top x) \mu (\ud x, \ud t).
\end{equation}
The last term $\int_0^t \int_{\Real^d} (\xi^\top x) \mu (\ud x, \ud
t)$ is a pure-jump increasing process, and since $\xi \in \I$ we
have $\xi^\top b - \int \xi^\top \hx \nu(\ud x) \geq 0$. Finally,
since $\xi \notin \N$ we must have that one of the two processes in
the right-hand-side of (\ref{eq: decomp of UIP}) is nonzero; it
follows that $\xi^\top X$ is increasing and nonzero, and thus $W^\xi
= \Exp(\xi^\top X)$ is increasing and nonconstant ($\prob[W^\xi_T >
1] > 0$).

Let us now assume that for some $\xi \in \Real^d$ we have $W^\xi$
being increasing; this is equivalent to saying that $\xi^\top X$ is
increasing. But then it is of finite variation, thus $\xi^\top \beta
= 0$, i.e., $\xi^\top c = 0$. Further, we must have $\xi^\top \Delta
X \geq 0$ which is of course equivalent to $\nu[\xi^\top x < 0] =
0$. Finally, we can write $\xi^\top X$ as in (\ref{eq: decomp of
UIP}) and since $\xi^\top X$ is increasing, the first term is
continuous (linear) and the second pure-jump we must have $\xi^\top
b - \int \xi^\top \hx \nu(\ud x) \geq 0$. We have all three
conditions of Definition \ref{dfn: instant_arb_opport}, and finally
if $\xi^\top X$ is nonzero we must have $\xi \notin \N$, which gives
$\xi \in \I$. \qed

\subsection{Changes of measure that respect the exponential L\'evy structure} \label{subsec: Levy change of measure}
Absence of free lunches in the market is connected to existence of
probability measures that are equivalent to the original and endow
the stock price processes with some martingale-type property. In the
context of exponential L\'evy models it is actually possible to
change the original probability $\prob$ in such a way so that the
exponential L\'evy property remains intact. We now describe a way of
doing so that will prove most useful in the proof of Theorem
\ref{thm: No Arbitrage for Levy finite time constrained}.

Pick $\eta \in \Real^d$ and then some $g : \Real^d \mapsto \Real$
such that $g(x) = 0$ for $|x | \leq 1$, as well as $\int e^{-
\eta^\top x - g(x)} \indic_{\{ |x| > 1 \}} \nu (\ud x) < + \infty$
--- for example, this will hold for every $\eta \in \Real^d$ if $g$
is defined by $g(x) = 0$ for $| x | \leq 1$ and $g(x) = | x|^2 - 1$
for $| x | > 1$ (this is exactly the function $g$ we shall use in
the sequel). The process $Z^{(\eta, g)}$ defined by
\begin{equation} \label{eq: Levy change of measure}
Z^{(\eta, g)}_t := \exp \Big( -\eta^\top X_t - \sum_{0 < s \leq t} g(\Delta X_s) - t \psi (\eta,
g) \Big),
\end{equation}
for some constant $\psi (\eta, g)$ is exponential L\'evy and the
exponential formula (\ref{eq: expo formula}) give us that $\psi
(\eta, g) := - \eta^\top b + \frac{1}{2} \eta^\top c \eta + \int
(e^{- \eta^\top x - g(x)} - 1 + \eta^\top \hx) \nu (\ud x)$ makes
$Z^{(\eta, g)}$ a martingale.

Define then a new probability measure $\prob^{(\eta, g)}$ via $(\ud
\prob^{(\eta, g)} / \ud \prob) |_{\F_T} = Z^{(\eta, g)}_T$. Pick any
positive Borel-measurable functional $\Phi$ that acts on c\'adl\'ag
processes and observe that for all $0 \leq t \leq T$ we have, with
$\expec^{(\eta, g)}$ denoting expectation under $\prob^{(\eta, g)}$:
\begin{eqnarray*}
  \expec^{(\eta, g)} \left[ \Phi \big( (X_{t + s} - X_t)_{0 \leq s \leq T-t} \big) \ \big| \ \F_t \right] &=& \expec \Big[ \frac{Z^{(\eta, g)}_T}{Z^{(\eta, g)}_t} \Phi \big( (X_{t + s} - X_t)_{0 \leq s \leq T-t} \big) \ \big| \ \F_t \Big] \ =\\
  \expec \big[ \widehat{\Phi} \big( (X_{t + s} - X_t)_{0 \leq s \leq T-t} \big) \ \big| \ \F_t \big]      &=& \expec \big[ \widehat{\Phi} \big( (X_s)_{0 \leq s \leq T-t} \big) \big] \ =  \\
  \expec [ Z^{(\eta, g)}_{T-t}  \Phi ( (X_s)_{0 \leq s \leq T-t} ) ]                      &=& \expec^{(\eta, g)} [\Phi ( (X_s)_{0 \leq s \leq T-t} ) ].
\end{eqnarray*}
The functional $\widehat{\Phi}$ above has obvious definition. It
follows that $X$ is still a L\'evy process under $\prob^{(\eta,
g)}$. Since $Z^{(\eta, g)}_t e^{i u^\top X_t} = \exp [ (i u -
\eta)^\top X_t - \sum_{0 < s \leq t} g(\Delta X_s) - t \psi (\eta,
g) ]$, we have
\[
\expec^{(\eta, g)} [e^{i u^\top X_t}] = \exp \big( t ( \psi(\eta - i
u, g) - \psi(\eta, g) ) \big);
\]
thus, the cumulant $\phi^{(\eta, g)}$ (the equivalent of (\ref{eq:
cummulant of Levy}) under the probability $\prob^{(\eta, g)}$)
satisfies $\phi^{(\eta, g)} (u) = \psi(\eta - i u, g) - \psi(\eta,
g)$. Straightforward computations give the L\'evy triplet
$(b^{(\eta,g)}, c^{(\eta,g)}, \nu^{(\eta,g)})$ of $X$ under
$\prob^{(\eta, g)}$ to be $b^{(\eta, g)} =  b - c \eta + \int (e^{-
\eta^\top x - g(x)} - 1) \hx \nu(\ud x)$, $c^{(\eta, g)} = c$ and
$\nu^{(\eta, g)} = e^{- \eta^\top x - g(x)} \nu (\ud x)$. Definition
\ref{dfn: instant_arb_opport}, coupled with the last equations
involving the L\'evy triplet of $X$ under $\prob^{(\eta, g)}$, imply
that \emph{the set $\I$ of immediate arbitrage opportunities remains
invariant} when we change from $\prob$ to $\prob^{(\eta, g)}$.

Let us finally remark that the transition from $\prob$ to
$\prob^{(\eta, g)}$ can be carried out in two steps. First, we
change $\prob$ to $\prob^{(0,g)}$ ``lightening'' the tails of the
L\'evy measure using the function $e^{-g}$, which turns out to be
exactly the Radon-Nikodym derivative of $\nu^{(0, g)}$ (the L\'evy
measure of $X$ under $\prob^{(0,g)}$) with respect $\nu$ (the L\'evy
measure of $X$ under $\prob$). As a second step, we change
$\prob^{(0, g)}$ to $\prob^{(\eta, g)}$, \emph{exponentially
tilting} $\prob^{(0, g)}$. This exponential tilting method is also
referred to as the \textsl{Esscher transform}.


\subsection{No-free-lunch equivalences for the cone-constrained case}

We are almost ready present a complete characterization of the
arbitrage situation in exponential L\'evy financial models for the
finite time-horizon case and a constrained set $\C$ that is a closed
convex \emph{cone} with $\N \subseteq \C$. There is one formal
definition missing involving the ability to change the original
measure $\prob$ to some other equivalent probability measure
$\qprob$ such that the stock price process, or possibly only the
allowed wealth processes $W^\pi$ for $\pi \in \Pi_\C$  have some
kind of martingale property under $\qprob$.

In the unconstrained case, the notion of an equivalent martingale
measure (see Definition \ref{dfn: equivalnet mart measure} below)
does the trick for our no-free-lunch equivalences, but in the
presence of constraints this is no longer the case. The reason is
that free lunches are not allowed only for portfolios that take
values in $\C$. Further, we cannot even hope that all \emph{wealth
processes} are martingales. Take for example $X$ to be the negative
of a Poisson process and assume we are constrained in the cone of
positive strategies $\C = \Real_+$. Under any measure $\qprob \sim
\prob$, the process $S = \Exp(X)$ will be non-increasing and not
identically equal zero, which prevents it from being (even a local)
martingale. It is a supermartingale though, and this turns out to be
the appropriate notion.

\begin{defn} \label{dfn: equivalnet mart measure}
A probability $\qprob$ that is equivalent to $\prob$ (we denote
$\qprob \sim \prob$) will be called

\noindent $\bullet$ \textsl{equivalent martingale measure} (EMM in
short) if the discounted stock-price $S$ is a vector
$\qprob$-martingale.

\noindent $\bullet$ \textsl{$\C$-constrained equivalent
supermartingale measure} (\ESMMC in short) if the wealth process
$W^\pi$ is a $\qprob$-supermartingale for all $\pi \in \Pi_\C$. The
class of all \ESMMC is denoted by $\QprobC$.
\end{defn}

Stochastic integrals of martingales that are further positive
processes are local martingales; this has been shown by Ansel and
Stricker \cite{Ansel-Stricker}. Further, it is well-known that
positive local martingales are supermartingales. Thus, we get that
an EMM a fortiori is an \ESMMC for any $\C$; of course the opposite
does not hold in general.

Even if $\C$ is just a convex set, it is easy to see that if an
\ESMMC exists then it is automatically an equivalent supermartingale
measure for the market with \emph{cone} constraints
$\overline{\cone} (\C)$, the \emph{closure of the smallest cone that
contains $\C$}; the proof of this simple statement is left to the
diligent reader. Thus, if we want to prove any theorem concerning
equivalent supermartingale measures we might as well assume cone
constraints
--- the pure convex case is treated in the next section.

\smallskip

For exponential L\'evy models, and even under the \emph{weakest} of
no-free-lunch conditions (namely, NUIP$_\C$), not only can we find
an ESMM$_\C$, but we can do so in a matter that respects the
exponential L\'evy structure as was described in the previous
subsection.

\begin{thm} \label{thm: No Arbitrage for Levy finite time constrained}
For an exponential L\'evy model with closed convex cone constraints
$\C$ on a finite financial planning horizon $[0,T]$, the following
are equivalent:
\begin{enumerate}
    \item There exists a $\qprob \sim \prob$ under which $X$ remains a L\'evy process and $\pi^\top X$ is a
    L\'evy supermartingale for all $\pi \in \C$.
    \item The \emph{ESMM$_\C$} condition holds: $\QprobC \neq \emptyset$;
    \item The \emph{NFLVR$_\C$} condition holds;
    \item The \emph{NA$_\C$} condition holds;
    \item The \emph{NUPBR$_\C$} condition holds;
    \item The \emph{NUIP$_\C$} condition holds;
    \item $\I \cap \C = \emptyset$.
\end{enumerate}
\end{thm}

\proof The implication (1) $\Rightarrow$ (2) is obvious: (1) is
stronger than (2).

For (2) $\Rightarrow$ (3), we have that $W^\pi$ for all $\pi \in
\Pi_\C$ is a positive $\qprob$-supermartingale. Consider a sequence
$(\pi_n)_{n \in \Natural}$ of elements in $\Pi_\C$ that is a
candidate for being a free lunch with vanishing risk, i.e., suppose
that there exists a sequence $(\delta_n)_{n \in \Natural}$ with
$\delta_n \downarrow 0$ and $\prob[W^{\pi_n}_T \geq 1 - \delta_n] =
1$. Then, for all $\epsilon
> 0$, $(1 + \epsilon) \qprob[W^{\pi_n}_T > 1 + \epsilon] + (1 - \delta_n)
(1 - \qprob[W^{\pi_n}_T > 1 + \epsilon]) \leq \expec^\qprob
W^{\pi_n}_T \leq 1$, which by simple algebra manipulations implies
$\qprob[W^{\pi_n}_T > 1 + \epsilon] \leq \delta_n / (\epsilon +
\delta_n)$. The right-hand-side of this last inequality converges to
zero as $n$ tends to infinity; since $\prob \sim \qprob$ we have
that $\lim_{n \to \infty} \prob[W^{\pi_ n}_T > 1 + \epsilon] = 0$ as
well, and NFLVR holds.

The implications (3) $\Rightarrow$ (4) and (3) $\Rightarrow$ (5) are
an easy exercise (use Definition \ref{dfn: arbitrage notions}), and
implications (4) $\Rightarrow$ (6) and (5) $\Rightarrow$ (6) are
even easier.

Implication (6) $\Rightarrow$ (7) is one direction of Lemma
\ref{lem: imm arb opp implies UIP}.

\smallskip

The cycle will be closed as soon as we prove (7) $\Rightarrow$ (1),
which is the harder one. As mentioned in the Introduction, we follow
the idea of Rogers \cite{Rogers}, who applied it for discrete-time
processes. Using the notation of the previous subsection
\ref{subsec: Levy change of measure}, begin by changing the measure
$\prob$ into $\prob^{(0,g)}$, where $g$ is defined by $g(x) = 0$ for
$| x | \leq 1$ and $g(x) = |x|^2 - 1$ for $| x | > 1$. Then
$\expec^{(0,g)} [\exp (| X_T |^2)] < \infty$; this is due to the
behavior of the tails of the L\'evy measure $\nu^{(0,g)}$ (in the
notation of subsection \ref{subsec: Levy change of measure}) under
$\prob^{(0,g)}$ --- one can check for example Sato \cite{Sato} for
matters like this. Since $X$ is still a L\'evy process under
$\prob^{(0,g)}$ and $\I$ remains invariant under this change of
measure we might as well assume from the outset that $\expec [\exp
(| X_T |^2)] < \infty$ (i.e., $\prob \equiv \prob^{(0,g)}$).

We proceed by considering the \textsl{exponential utility} function
$U(x) := 1 - e^{-x}$ and setting $\phi(\p) := \expec U(\p^\top X_T)
= 1 - \expec [e^{-\p^\top X_T}]$. The function $\phi$ is real-valued
(because $\expec \exp (| X_T |^2) < \infty$) and concave. Let
$\phi_* := \sup_{\p \in \C} \phi(\p)$; since $\phi(\p) = \phi(\p +
\zeta)$ for $\zeta \in \N$, nothing changes if we restrict this
infimum on $\N^\bot$ (see Remark \ref{rem: does not hurt to consider
non-degeneracy}). Clearly, $\phi_* \geq \phi(0) = 0$.

We claim that if $\I = \emptyset$, the supremum $\phi_*$ is achieved
by a point in $\N^\bot \cap \C$. Otherwise, there would exist a
sequence $(\p_n)_{n \in \Natural}$ in $\N^\bot \cap \C$ such that
$\lim_{n \to \infty} \uparrow | \p_n | = + \infty$, $\phi(\p_n) \in
\Real_+$ and $\lim_{n \to \infty} \phi(\p_n) = \phi_*$. Then, set
$\xi_n := \p_n / | \p_n |$ and fix $a \in \Real_+$; eventually, for
all $n \geq n_a$ where $n_a$ is large enough to satisfy $a \leq |
\p_{n_a} |$, we have $a \xi_n \in \N^\bot \cap \C$ and $\phi(a
\xi_n) \geq 0$ (the last follows from concavity of $\phi$ as soon as
one remembers that $\phi(0) = 0$ and $\phi(\p_n) \geq 0$). Since
$(\xi_n)_{n \in \Natural}$ is a sequence of unit vectors in $\N^\bot
\cap \C$ we can assume without loss of generality that it converges
to some unit vector $\xi \in \N^\bot \cap \C$ (choosing a
subsequence otherwise). Since $U (x) \leq 1$ for all $x \in \Real$,
Fatou's lemma is applicable and will give
\[
\phi(a \xi) = \expec U (a \xi^\top X_T) \geq \limsup_{n \to \infty}
\expec U (a \xi_n^\top X_T) = \limsup_{n \to \infty} \phi(a \xi_n)
\geq 0.
\]
In other words, $\expec [(e^{- \xi^\top X_T})^a] \leq 1$ for all $a
\in \Real_+$; this can only hold if $\prob[\xi^\top X_T \geq 0] =
1$. Since $\xi^\top X$ is a L\'evy process, Lemma \ref{lem: pos is
incr for Levy} suggests that $\xi^\top X$ is increasing; since $\xi
\in \N^\bot$, Lemma \ref{lem: imm arb opp implies UIP} would finally
give $\xi \in \I \cap \C$, which is assumed empty. We reached a
contradiction to our assumption because we assumed that the supremum
of $\phi$ is not attained by any vector in $\N^\bot \cap \C$. Thus,
there exists $\eta \in \N^\bot \cap \C$ such that $\phi(\eta) =
\phi_*$.

Now, pick any $\p \in \C$ and observe that $\Real_+ \ni a \mapsto
\phi (\eta + a \p)$ is concave in $a \in \Real_+$ that has a maximum
at $a = 0$. It follows that
\[
\expec \Big[ \frac{e^{-\eta^\top X_T} - e^{- (\eta + a \p)^\top
X_T}}{a} \Big] = \frac{\phi(\eta + a \p) - \phi(\eta)}{a} \leq 0,
\textrm{ for all } a > 0.
\]
The concavity of $x \mapsto e^{-x}$ implies that the expression
inside the expectation above is an increasing function of decreasing
$a$; it is also clear that it converges $\prob$-a.s. to
$e^{-\eta^\top X_T} \p^\top X_T$ as $a \downarrow 0$. Since $\phi$
is finite-valued, we can use the monotone convergence theorem to get
$\expec[e^{-\eta^\top X_T} \p^\top X_T] \leq 0$. In other words,
defining $\prob^{(\eta, 0)}$ as in subsection \ref{subsec: Levy
change of measure} we get $\expec^{(\eta, 0)} [\p^\top X_T] \leq 0$
for all $\p \in \C$. This means that $\p^\top X$ is a L\'evy
supermartingale for all $\p \in \C$. \qed

\begin{rem} (\textsc{On the Unconstrained Case}) Recall from Remark \ref{rem: on natural constraints}
the natural constraints set $\C_0$. Then If $\C_0 \subseteq \C$,
i.e., in the unconstrained case, then one can replace conditions (1)
and (2) of Theorem \ref{thm: No Arbitrage for Levy finite time
constrained} above by

(1') \emph{There exists $\qprob \sim \prob$ under which $X$ is
L\'evy martingale and $S$ martingale}.

(2') \emph{An \EMM exists};

\smallskip

\noindent Indeed (1') $\Rightarrow$ (2') is obvious, while (1)
$\Rightarrow$ (1') follows like this: $\p^\top X$ being a
$\qprob$-martingale for all $\p \in \Real^d$ means that $X$ is a
$\qprob$-martingale. Then, each $S^i$, $i = 1, \ldots, d$ is a
positive local martingale; the exponential formula (\ref{eq: expo
formula}) gives $\expec^{\qprob} S_T = S_0$, i.e., that $S$ is a
martingale.
\end{rem}

\begin{rem} (\textsc{Martingale vs $\sigma$-Martingale Measures})
In their seminar work, Delbaen and Schachermayer \cite{DS: FTAP
unbounded} have showed that in a general semimartingale model in the
unconstrained case and a possibly non locally bounded asset-price
process $S$, the NFLVR condition is equivalent to existence of some
$\qprob \sim \prob$ such that $S$ is a \textsl{$\sigma$-martingale}
under $\qprob$ (which basically means that we can write $S$ as a
stochastic integral of a martingale).

For exponential L\'evy markets, it turns out from the previous
remark that  \emph{any} of our no-free-lunch conditions is
equivalent to the existence of an EMM. There has been work from some
authors (we mention for example Cherny \cite{Cherny : NGA} and Yan
\cite{Yan}) on establishing a version of the FTAP in which
no-free-lunch criteria are equivalent to the existence of an EMM,
instead of simple a $\sigma$-martingale one. Obviously, these
no-free-lunch criteria are equivalent to the ones mentioned in
Theorem \ref{thm: No Arbitrage for Levy finite time constrained}. In
particular, Yan's work \cite{Yan} allows us to conclude that we can
\emph{enlarge} the class of strategies that agents can use. Indeed,
any predictable process $\theta$ (where now $\theta^i_t$ is
perceived as the \emph{units} of asset $i$ that is held by the agent
at time $t$) such that $\theta \cdot S \geq - a(1 + \sum_{i=1}^d
S^i)$ for some $a > 0$ is allowed, and will not lead to free lunch.
\end{rem}

\begin{rem} \label{rem: minimal rel entopy and exp util maxim} (\textsc{On Exponential Utility Maximization})
The \ESMMC $\qprob$ in the proof of equivalence (7) $\Rightarrow$
(1) in Theorem \ref{thm: No Arbitrage for Levy finite time
constrained} above is constructed via \emph{exponential utility
maximization} in the
financial market where 
the ``original'' probability measure is $\prob^{(0,g)}$. We are not
able to use directly $\prob$ because $\expec[e^{\p^\top X_T}]$ might
be infinite for some $\p \in \C$; in case $\expec[e^{\p^\top X_T}] <
\infty$ for all $\p \in \C$ we can proceed with the proof and the
measure $\qprob = \prob^{(\eta, 0)}$ that we end up with is the
\textsl{minimal entropy martingale measure}. The theme has received
a lot of attention, let us just mention here that it has been
treated by Fujiwara and Miyahara \cite{Fuj-Miya} and recently by
Esche and Schweizer \cite{Esc-Schw}, as well as Hubalek and Sgarra
\cite{Hub-Sga}.

Nevertheless, if $\expec[e^{\p^\top X_T}]$ could take possibly
infinite values, things are slightly more complicated. In that case,
we can still find a vector $\eta \in \C$ such that $\expec
[U(\eta^\top X_T)] \geq \expec [U(\p^\top X_T)]$ for all $\p \in \C$
(under the assumption $\I = \emptyset$, of course), but \emph{we
cannot conclude that $\prob^{(\eta, 0)}$ is an equivalent martingale
measure}. Take for example a unconstrained, one-stock exponential
L\'evy model with $c = 0$ and L\'evy measure $\nu$ of the form
$\nu(\ud x) = f(x) \ud x$ with $f(x) > 0$ for all $x \geq 1$ (so
that $\I = \emptyset$), and (i) $\int e^{a x} \indic_{ \{x
> 1 \}} f(x) \ud x = \infty$ for all $a
> 0$, (ii) $\int x \indic_{\{x > 1\}} f(x) \ud x < \infty$, and
(iii) $b + \int x \indic_{\{x > 1\}} f(x) \ud x < 0$. An example of
such density $f$ satisfies $f(x) \sim x^{-p}$ as $x \to \infty$ for
some $p > 2$; then (i) and (ii) hold automatically and an
appropriate choice of small enough $b$ will ensure (iii) as well.
Now, with $\phi(\p) := 1 - \expec e^{-\p X_T}$ we have $\phi(\p) = -
\infty$ for all $\p < 0$, and a simple use of Jensen's inequality
gives $\phi(\p) < 0 = \phi(0)$ for all $\p > 0$ (because by (iii) we
have $\expec [\p X_T] < 0$ for $\p
> 0$). It follows that the optimal portfolio is $\eta = 0$; this
gives us $\qprob = \prob$, which is \emph{not} an equivalent
martingale measure, since $\expec X_T < 0$ by (iii). Observe
nevertheless that it is an ESMM, and it can be shown that it will
always be
--- this is not just a coincidence here.
\end{rem}

\subsection{Completeness} \label{subsec: completeness} Though not
our main concern, we give here a characterization of
\textsl{completeness} (the ability to perfectly replicate any
bounded contingent claim) in exponential L\'evy markets. We do not
provide full details --- we trust they can be filled by the reader.
We note however that the weak martingale representation property for
the filtration generated by a L\'evy process as described for
example in Jacod and Shiryaev \cite{Jacod - Shiryaev} will have to
be used.

\begin{defn} \label{dfn: completeness}
The exponential L\'evy market in a finite time-horizon $[0, T]$ is
called \textsl{complete} if for all positive and bounded $H \in
\F_T$ one can find $\pi \in \Pi$ and $x > 0$ such that $x W^\pi_T =
H$.
\end{defn}

In order to talk about completeness one should better assume that we
are in the unconstrained case $\C = \Real^d$ (thus the absence of a
subscript from $\Pi$ in the definition above), and that \emph{the
filtration $\mathbf{F}$ is the usual augmentation of the one
generated by $S$}, or equivalently of the one generated by $X$.
These conditions are in force for this subsection.

We decompose $\Real^d = \K \oplus \K^\bot$, where $\K := \{ x \in
\Real^d \such c x = 0\}$ is the \textsl{kernel} of the covariance
matrix $c$ and $\K^\bot$ is its orthogonal complement, and we denote
by $k$ the dimension of the linear subspace $\K$. We also denote by
$\supp (\nu)$ the \textsl{support} of the measure $\nu$, i.e., the
smallest closed subset of $\Real^d$ that $\nu$ gives full measure.

\begin{prop}
With the assumptions and notation set above (in particular, $\C =
\Real^d$) and an exponential L\'evy market on a finite time-horizon
$[0,T]$, suppose that the model satisfies any (and thus all) of the
equivalent conditions of Theorem \ref{thm: No Arbitrage for Levy
finite time constrained}. The following are equivalent:

\noindent \emph{(1)} The exponential L\'evy model is complete.

\noindent \emph{(2)} There exists a \emph{unique} \EMM $\qprob$.

\noindent \emph{(3)} We have (i) $\supp(\nu) \subseteq \K$, (ii)
$\supp(\nu)$ contains at most $k$ points.
\end{prop}

One can start directly from the exponential L\'evy model and not
assume that it satisfies the equivalent conditions of Theorem
\ref{thm: No Arbitrage for Levy finite time constrained}. In that
case, (1) should be substituted with

\noindent (1') The exponential L\'evy model satisfies any of the
conditions of Theorem \ref{thm: No Arbitrage for Levy finite time
constrained} \emph{and} is complete. \\ Implication (2) remains the
same, while for (3) we have to add an extra requirement (3 iii)
appearing below. To prepare the ground, notice that if (3) holds,
and with $X^\K$ denoting the orthogonal projection of $X$ on $\K$,
we have $X^\K_t = a t + \sum_{n = 1}^{N_t} Y_n$, for $a \in \K$, $N$
a Poisson process with some arrival rate $\lambda > 0$, and
$(Y_n)_{n \in \Natural}$ a sequence of i.i.d. (and independent of
$N$) random variables with simple discrete distributions charging
less than $k$ points on $\K$. Condition $\I = \emptyset$ of Theorem
\ref{thm: No Arbitrage for Levy finite time constrained} is now
equivalent to the following:

\noindent (3 iii) if $\xi \in \K$ satisfies $\xi^\top a \geq 0$ and
$\xi^\top x \geq 0$ for all $x \in \supp(\nu)$, then we actually
have $\xi^\top a = 0$ and $\xi^\top x = 0$ for all $x \in
\supp(\nu)$.

\section{The Num\'eraire Portfolio, Supermartingale Deflators and No-Free-Lunch Equivalences for Convex-Constrained Models} \label{sec: Numeraire portfolio}

In this section we aim in extending the scope of Theorem \ref{thm:
No Arbitrage for Levy finite time constrained} to the
convex-constrained case. As a byproduct we shall obtain even more
equivalences for the cone-constrained and unconstrained case then
the ones covered by Theorem \ref{thm: No Arbitrage for Levy finite
time constrained}. We introduce a very special portfolio that will
help us do that. As discussed in Remark \ref{rem: minimal rel entopy
and exp util maxim}, in the course of proving Theorem \ref{thm: No
Arbitrage for Levy finite time constrained} we used the optimal
portfolio for \emph{exponential} utility for a possibly changed
probability measure; vis-\`a-vis, here we shall use the optimal
portfolio for \emph{logarithmic} utility under the \emph{original}
measure $\prob$. This will enable us to prove equivalences valid
under closed and convex
--- but not necessarily cone --- constraints; more importantly, it
is exactly \emph{this} result that allows for generalization in
general semimartingale models. The drawback is that we have to work
harder; part of the proof of the main result here (Theorem \ref{thm:
num iff def-non-empty iff NUPBR iff NUIP}) is more technical and
long, and will be the focus of the next section --- this contrasts
the (fair) easiness of the proof of Theorem \ref{thm: No Arbitrage
for Levy finite time constrained}. After the work is done, we
continue the story in Karatzas and Kardaras \cite{KK: num and
arbitrage} for the semimartingale case.

\subsection{The inadequacy of equivalent supermartingale measures}

As soon as we face \emph{non-conic} convex constraints, the \NAC ---
or even \NFLVRC --- condition is not any more sufficient to imply
existence of an equivalent supermartingale measure.

\begin{ex} \label{ex: NA does not imply ESMM}
We take $X$ be a 2-dimensional compound Poisson process, i.e., $X_t
= \sum_{i=1}^{N_t} Y_i$, for $t \in [0, T]$, where $N$ is a standard
Poisson process and $Y_i$ is a sequence of 2-dimensional independent
and identically distributed random variables with $Y_i = (e_i, f_i -
1)$, $e_i$ and $f_i$ being independent with a standard exponential
distribution (we only use the fact that they are independent and
their distributions are supported on the positive half-line --- even
less is needed as the reader will note). Of course, in the
unconstrained case there is clear arbitrage: take a strict long
position in the first stock and null position on the second.
Consider now the constraints set $\C := \{(x,y) \in \Real^2 \such
x^2 \leq y \}$, i.e., only points on and above the parabola $y =
x^2$ are allowed for investing. We claim that \NFLVRC holds, but no
\ESMMC exists.

To see that no \ESMMC exists is easy: we have already noted that if
it did it should already be an equivalent supermartingale measure
for the market with constrains $\overline{\cone}(\C) = \Real_+
\times \Real$; the latter is clearly impossible, since there is
arbitrage.

In the process of showing the no free lunches exist for the
$\C$-constrained market, we use the following observation: for $\p
\equiv (x,y) \in \C \setminus \{0\}$ it \emph{must} be that $y > 0$
(due to the constraints $y = 0 \Rightarrow x = 0$); also, since
$\prob[e_1 > 0] = 1$, we have $x e_1 + y (f_1 - 1) \leq \sqrt{y} e_1
+ y (f_1 - 1)$. Then, $\prob[\p^\top \Delta Y_1 < 0] \geq \prob[e_1
< \sqrt{y}(1-f_1)]
> 0$; this should already give you a hint why no $\C$-constrained arbitrage exists.

We now show that \NAC holds. Pick any portfolio $\pi \in \Pi_\C$
that is supposed to generate an arbitrage and define $\tau := \inf
\{t \in [0, T] \such \Delta W^\pi_t \neq 0 \}$, where we set $\tau =
T$ when the set that we are taking the infimum is empty. It is
obvious that $\tau$ is an $\mathbf{F}$-stopping time; actually, with
$\tau_n := \inf \{t \in \Real_+ \such N_t = n\}$ denoting the
n$^\textrm{th}$ jump of $N$, we have $\{ \tau = \tau_n\} =
\{\pi_{\tau_k}=0 \textrm{ for all } k < n, \ \pi_{\tau_n} \neq 0\}
\in \F_{\tau_n -}$, a fact that will be important. Now, $\{\tau = T
\} \subseteq \{W^\pi_T = 1\}$, thus if $\prob[\tau = T] = 1$ we have
$\prob [W^\pi_T = 1] = 1$ and $\pi$ is not an arbitrage. Suppose
then that $\prob[\tau < T] > 0$; we shall show that $\prob[W^\pi_T <
1, \ \tau < T] > 0$, and then \NAC readily follows. Define the
\emph{second} time that a wealth readjustment happens $\tau' := \inf
\{t \in (\tau, T] \such \Delta W^\pi_t \neq 0 \}$, where again we
set $\tau' = T$ if the last set is empty. $\tau'$ is an
$\mathbf{F}$-stopping time and we have $\prob[W^\pi_T < 1] \geq
\prob[W^\pi_T < 1, \ \tau < T, \ \tau' = T] = \prob[\pi^\top_{\tau}
\Delta X_{\tau} < 0, \ \tau < T, \ \tau' = T]$. Since $\{ \tau =
\tau_n\} \in \F_{\tau_n -}$, Lemmata \ref{lem: jump indep of strict
hist} and \ref{lem: sampled jump has same distr as first} in the
Appendix give that $\pi_{\tau} \in \F_{\tau -}$ is independent of
$\Delta X_{\tau}$ and that the latter jump is distributed as $Y_1$.
On $\{\tau < T \}$ we have $\pi_{\tau} \in \C \setminus \{0\}$; the
observation made in the previous paragraph coupled with the trivial
fact $\prob[\tau < T, \ \tau' = T] > 0$ imply $\prob[\pi^\top_{\tau}
\Delta X_{\tau} < 0, \ \tau < T, \ \tau' = T]
> 0$, and thus $\prob[W^\pi_T < 1] > 0$. We conclude that \NAC holds
for this constrained market.

The fact that \NAC holds implies that actually \NFLVRC holds as
well. The reason is that finite-time-horizon
compound-Poisson-process models are equivalent to discrete-time
models with a stochastic, but \emph{finite} time-horizon; for
discrete-time models, it is not hard to see that \NFLVRC is
equivalent to the generally weaker \NAC (this is no longer true for
infinite time-horizon models).
\end{ex}

\subsection{The \num portfolio}

The following concept will prove crucial.

\begin{defn} \label{dfn: numeraire}
A portfolio $\rho \in \Pi_\C$ will be called \textsl{\num portfolio}
for the class $\Pi_\C$, if for every other $\pi \in \Pi_\C$ the
\textsl{relative wealth process} $W^\pi / W^\rho$ is a
supermartingale.
\end{defn}

The reader is referred to in Becherer \cite{Becherer} for the
definition and more on this concept. The \num portfolio has many
optimality properties; you can check Karatzas and Kardaras \cite{KK:
num and arbitrage}, where an extensive discussion on the existence
of the \num portfolio for general semimartingale models and its
relationship with free lunches is taking place.

\begin{ex} \label{ex: zero num iff P is supermart measure}
The \num portfolio exists and is equal to zero if and only if all
wealth processes $W^\pi$ for $\pi \in \Pi_\C$ are
$\prob$-supermartingales. This is a trivial example, but it will
find use in Theorem \ref{thm: No Arbitrage for inf-horizon Levy}
where arbitrage in infinite-time horizon exponential L\'evy markets
is studied.
\end{ex}

\subsection{Equivalent supermartingale deflators}

We introduce a concept that is weaker --- but very closely
related --- to equivalent supermartingale measures. Let us assume
that the \num portfolio exists; by way of definition, the process
$\pare{W^\rho}^{-1}$ acts as a ``deflator'', under which all wealth
processes $W^\pi$ for $\pi \in \Pi_\C$ become supermartingales.
There are more processes sharing this last property.

\begin{defn} \label{dfn: supermartingale deflators}
A process $D$ will be called a \textsl{$\C$-constrained equivalent supermartingale deflator}
(ESMD$_\C$) if $D_0 = 1$, $D_T > 0$ and such that
$D W^\pi$ is a supermartingale for all $\pi \in \Pi_\C$. The class
of all ESMD$_\C$'s is denoted by $\Def_\C$.
\end{defn}

A \ESMMC (say, $\qprob$) generates an \ESMDC $D$ via the density process $D_t = \pare{\ud \qprob / \ud \prob} |_{\F_t}$, for $t
\in [0, T]$, so that $\QprobC \neq \emptyset \Rightarrow \Def_\C
\neq \emptyset$. The reverse implication $\Def_\C \neq \emptyset
\Rightarrow \QprobC \neq \emptyset$ does not hold in general as a
simple example involving the notorious three-dimensional Bessel
process shows; see Delbaen and Schachermayer \cite{DS: Bessel}.
Nevertheless, for exponential L\'evy models and under \emph{cone}
constraints we shall soon see that $\Def_\C \neq \emptyset
\Rightarrow \QprobC \neq \emptyset$ \emph{does} hold.

\subsection{The main result}

Here is the result that puts the \num portfolio in the context of
arbitrage. The difficult implication below is (5) $\Rightarrow$ (1)
and will be the result of discussion in the subsequent subsections
and the following section \ref{sec: crucial lemma}.

\begin{thm} \label{thm: num iff def-non-empty iff NUPBR iff NUIP}
For an exponential L\'evy model under closed convex constraints $\C
\subseteq \Real^d$ on a finite-time horizon $[0, T]$, the following
are equivalent:
\begin{enumerate}
    \item The \num portfolio exists in the class $\Pi_\C$.
    \item An \emph{ESMD$_\C$} exists: $\Def_\C \neq \emptyset$.
    \item The \emph{NUPBR$_\C$} condition holds.
    \item The \emph{NUIP$_\C$} condition holds.
    \item $\I \cap \check{\C} = \emptyset$.
\end{enumerate}
If $\C$ is further a cone ($\C = \check{\C}$), (1) and (2) above are
equivalent to all conditions of Theorem \ref{thm: No Arbitrage for
Levy finite time constrained}.
\end{thm}

\proof The implication (1) $\Rightarrow$ (2) is trivial:
$(W^\rho)^{-1}$ is an element of $\Def_\C$.

Now, for the implication (2) $\Rightarrow$ (3), start by assuming
that $\Def_\C \neq \emptyset$ and pick an element $D \in \Def_\C$ .
We wish to show that $\{ W^\pi_T \such \pi \in \Pi_\C \}$ is bounded
in probability. Since $D_T >0$, this is equivalent to showing that
$\{ D_T W^\pi_T \such \pi \in \Pi_\C \}$ is bounded in probability.
This easily follows from the fact that $D W^\pi$ for $\pi \in
\Pi_\C$ are positive supermartingales with $D_0 W^\pi_0 = 1$ and so,
for all $m > 0$, $\sup_{\pi \in \Pi_\C} \prob [ D_T W^\pi_T > m ]
\leq m^{-1} \sup_{\pi \in \Pi_\C} \expec[D_T W^\pi_T] \leq m^{-1}$.

The implication (3) $\Rightarrow$ (4) is (as already noticed)
trivial.

For (4) $\Rightarrow$ (5), if $\I \cap \check{\C} \neq \emptyset$
then Lemma \ref{lem: imm arb opp implies UIP} shows that \NUIPC
fails.

The implication (5) $\Rightarrow$ (1) is significantly harder; after
some preparation in the sequel, its proof will be the context of
Lemma \ref{lem: Levy charact of num} in the next section.

Finally, the claim for the further equivalences in the
cone-constrained case is obvious. \qed

\begin{rem}
Unless $\C$ is a cone, the conditions of Theorem \ref{thm: num iff
def-non-empty iff NUPBR iff NUIP} are \emph{not} equivalent to \NAC
in general. Actually, an increasing (but not unbounded) profit might
exist. Indeed, in the context of Example \ref{ex: NA does not imply
ESMM} consider the constraints set $\C = [0,1] \times [0,1]$. Since
$\Cc = \{0\}$, \NUIPC trivially holds, but of course $\pi = (1, 0)
\in \C$ is an increasing profit.
\end{rem}

\subsection{No-free-lunch equivalences in the infinite-time horizon case} \label{subsec: structure of pred convex constraints}

The situation for infinite-time horizon exponential L\'evy models is
drastically different than what we have seen in Theorems \ref{thm:
No Arbitrage for Levy finite time constrained} and \ref{thm: num iff
def-non-empty iff NUPBR iff NUIP}. It turns out that we can
\emph{always} construct free lunches (albeit not increasing profit
necessarily) unless the \emph{original} measure $\prob$ is
supermartingale measure, meaning that $W^\pi$ is a
$\prob$-supermartingale for all $\pi \in \Pi_\C$.

Previous definitions on free lunches, equivalent (super)martingale
measures and deflators can be read for infinite-time horizons by
plugging $T = + \infty$; the terminal wealths $W^\pi_T$ in
Definition \ref{dfn: arbitrage notions} have to be replaced by
$W^\pi_\infty = \lim_{t \to \infty} W^\pi_t$, where we assume that
this last limit exists $\prob$-a.s. (this is for example the case
when $\pi$ is supported on a stochastic interval $\dbra{0, \tau}$,
where $\tau$ is a $\prob$-a.s. finite stopping time).

\begin{thm} \label{thm: No Arbitrage for inf-horizon Levy}
For an exponential L\'evy stock-price model under closed convex
constraints $\C \subseteq \Real^d$ on a infinite-time horizon, the
following are equivalent:
\begin{enumerate}
    \item $W^\pi$ is a $\prob$-supermartingale for all $\pi \in \Pi_\C$.
    \item An \emph{ESMM}$_\C$ exists: $\QprobC \neq \emptyset$;
    \item An \emph{ESMD}$_\C$ exists: $\Def_\C \neq \emptyset$;
    \item The \emph{NFLVR}$_\C$ condition holds;
    \item The \emph{NUPBR}$_\C$ condition holds;
    \item The \emph{NA}$_\C$ condition holds.
\end{enumerate}
\end{thm}

\begin{rem} \label{rem: on NA for inf-hor levy and pred char}
Even though there is no \emph{direct} reference to a condition
involving the L\'evy triplet $(b,c,\nu)$ as there was in Theorems
\ref{thm: No Arbitrage for Levy finite time constrained} and
\ref{thm: num iff def-non-empty iff NUPBR iff NUIP} for finite-time
horizons, observe that actually condition (1) of Theorem \ref{thm:
No Arbitrage for inf-horizon Levy} is one. Indeed, in order for
$\prob$ to be such that $W^\pi$ is a $\prob$-supermartingale for all
$\pi \in \Pi_\C$ it is necessary and sufficient that $\expec[\p^\top
X_1] \leq 0$ (this does not mean that $\p^\top X_1$ is integrable
--- just that the positive part is integrable) for all $\p \in \C
\cap \C_0$.  In other words, for every $\p \in \C \cap \C_0$ we must
have $\p^\top b + \int \p^\top \hbarx \nu (\ud x) \leq 0$.
\end{rem}

\proof The implications (1) $\Rightarrow$ (2) $\Rightarrow$ (3)
$\Rightarrow$ (4) $\Rightarrow$ (5) and (4) $\Rightarrow$ (6) are
all trivial. We only prove (5) $\Rightarrow$ (1) and (6)
$\Rightarrow$ (1) below by showing that if $\prob$ is not a
supermartingale measure, both \NUPBRC and \NAC fail.

Assume then that $\prob$ is not a supermartingale measure. If $\I
\cap \check{\C} \neq \emptyset$, then \NUIPC fails and so both
\NUPBRC and \NAC will fail. On the other hand, if $\I \cap
\check{\C} = \emptyset$, the \num portfolio exists: it is a constant
portfolio $\rho$ that gives rise to a positive supermartingale
$(W^\rho)^{-1}$. We know that $(W^\rho_\infty)^{-1} := \lim_{t \to
\infty} (W^\rho_t)^{-1}$ exists $\prob$-a.s. in $\Real_+$. We
actually claim that $(W^\rho_\infty)^{-1} = 0$. Indeed, the fact
that this limit is a constant follows from Kolmogorov's 0-1 law for
the L\'evy process $L^\rho := \log W^\rho$; but we can only have
$L^\rho_\infty = + \infty$, for otherwise  $L^\rho$ would be a
L\'evy process with finite limit at infinity, which cannot happen
unless it is identically constant zero, and this would mean $W^\rho
\equiv 1$, or $\rho \in \N$ which cannot happen unless $\prob$ is a
supermartingale measure (see Example \ref{ex: zero num iff P is
supermart measure}) and we are working under the assumption that it
is not. Now, the fact $W^\rho_\infty = \infty$ allows us to
construct portfolios $\pi_n \in \Pi_\C$ by requiring $\pi_n := \rho
\indic_{\dbra{0, \tau_n}}$, where $\tau_n$ is the finite stopping
time $\tau_n := \inf \{ t \in \Real_+ \such W^\rho_t \geq n \}$.
Then, $W^{\pi_n}_\infty \geq n$ and both conditions \NUPBRC and \NAC
fail. \qed

\begin{rem} (\textsc{On the One-Dimensional, Unconstrained Case}).
For the infinite time-horizon case, Selivanov \cite{Selivanov} shows
that if $d=1$ and $\C = \Real^d$, then NFLVR is equivalent to the
following: either (1) $S$ is a $\prob$-martingale, or (2) $S$ is a
$\prob$-supermartingale and the jumps of $S$ are locally unbounded
above. We can actually get this result from Theorem \ref{thm: No
Arbitrage for inf-horizon Levy}: if the jumps of $S$ are locally
bounded above (equivalently, the jumps of $X$ are bounded above) we
have that $0$ belongs to the relative interior of the natural
constraints $\C_0$. From Remark \ref{rem: on NA for inf-hor levy and
pred char} this would mean that both $\expec[X_1] \leq 0$ and
$\expec[ - X_1] \leq 0$, which means that $X$, and thus $S$, is a
$\prob$-martingale.
\end{rem}

\subsection{Relative rate of return}
\label{subsec: nec+suf for a port to be the num}

In order to figure out whether a \emph{constant} vector $\rho \in
\C$ is the \num portfolio we should (at least) check that $W^\pi /
W^\rho$ is a supermartingale for all other constant $\pi \in \C$.
This is seemingly weaker than the requirement of Definition
\ref{dfn: numeraire}, but the two will actually turn out to be
equivalent.

Since for all $\pi$ and $\rho$ vectors in $\C$ we have that $W^\pi$
and $W^\rho$ are exponential L\'evy process we get that the
log-relative-wealth-process $L^{\pi | \rho} := \log (W^\pi /
W^\rho)$ is a L\'evy process itself. The exponential formula
(\ref{eq: expo formula}) implies that $\expec [W^\pi_T / W^\rho_T] =
\expec \exp (L^{\pi | \rho}_T) = \exp \big( T \rel (\pi \such \rho)
\big)$, where straightforward computations lead us to set
\begin{equation} \label{eq: rel_perf}
\rel(\pi \such \rho) \ := \ (\pi - \rho)^\top b - (\pi - \rho)^\top
c \rho + \int \bra{ \frac{(\pi - \rho)^\top x}{1 + \rho^\top x} -
(\pi - \rho)^\top \hx } \nu (\ud x).
\end{equation}
The quantity $\rel(\pi \such \rho)$ is the \textsl{relative rate of
return} of $\pi$ with respect to $\rho$.

The integrand appearing in (\ref{eq: rel_perf}) is equal to $(1 +
\pi^\top x)/(1 + \rho^\top x) - 1 - (\pi - \rho)^\top \hx$; this
quantity is bounded from below by $-1$ on $\{ |x| > 1 \}$ for the
L\'evy measure $\nu$, while on $\{ |x| \leq 1 \}$ behaves like
$(\rho - \pi)^\top x x^\top \rho$, which is comparable to $|x|^2$.
It follows that the integral always makes sense, but can take the
value $+ \infty$. In any case, the quantity $\rel(\pi \such \rho)$
of (\ref{eq: rel_perf}) is well-defined.

The relative wealth process $W^\pi / W^\rho$ is a supermartingale if
and only if $\expec[W^\pi_T / W^\rho_T] \leq 1$, equivalently if
$\rel(\pi \such \rho) \leq 0$. We remark that this result extends to
the case where $\pi$ (and $\rho$) are non-constant predictable
processes in $\Pi_\C$; the reason being that the predictable finite
variation part of $W^\pi / W^\rho = \exp(L^{\pi | \rho})$ --- given
that it is a special semimartingale and admits a Doob-Meyer
decomposition --- is $\int_0^\cdot \exp(L_{t-}^{\pi | \rho})
\rel(\pi_t \such \rho_t) \ud t$, one can check this directly or
refer to Karatzas and Kardaras \cite{KK: num and arbitrage}. The
previous discussion proves the following.

\begin{lem} \label{lem: necess and suff for numeraire}
In order for a constant vector $\rho \in \C$ to be the \num
portfolio in the class $\Pi_\C$ it is necessary and sufficient that
$\rel(\pi \such \rho) \leq 0$ for every $\pi \in \C$.
\end{lem}

It follows then that in order to prove the implication (5)
$\Rightarrow$ (1) in Theorem \ref{thm: num iff def-non-empty iff
NUPBR iff NUIP} it suffices to show that $\I \cap \check{\C} =
\emptyset$ implies that there exists a $\rho \in \C$ such that
$\rel(\pi \such \rho) \leq 0$ for every $\pi \in \C$; this is taken
on in Lemma \ref{lem: Levy charact of num}.

\subsection{The growth-optimal portfolio} \label{subsec: growth-optimal portfolios}

In this subsection we continue towards the goal to construct the
\num portfolio via the L\'evy triplet $(b,c,\nu)$ in case $\I \cap
\check{\C} = \emptyset$, using the fact that it is
\emph{essentially} equal to the growth-optimal portfolio, which has
been studied in Algoet and Cover \cite{Algoet-Cover} in a general
discrete-time setting. Take a constant portfolio $\pi \in \Pi_\C$;
its \textsl{growth rate} is defined as the drift rate of the
log-wealth process $\log W^\pi$. Since $\log W^\pi$ is a L\'evy
process, one can use (\ref{eq: lin formula}) and formally (since it
will not always exist) compute the growth rate of $\pi$ to be
\begin{equation} \label{eq: growth rate}
\g(\pi) := \pi^\top b - \frac{1}{2} \pi^\top c \pi + \int \bra{\log(1 + \pi^\top x) - \pi^\top
\hx} \nu (\ud x).
\end{equation}

It turns out that the \num portfolio and the \textsl{growth-optimal
portfolio} (defined as the one that maximizes the growth rate
(\ref{eq: growth rate}) over all portfolios) are essentially the
same.

\begin{ex} \label{ex: (NUIP) for continuous}
We consider the Samuelson-Black-Scholes-Merton model $X_t = b t +
\sigma \beta_t$, in the unconstrained case $\C = \Real^d$. According
to Example \ref{ex: special cases of imed arb opport} we have $\I =
\emptyset$ if and only if there exists $\rho \in \Real^d$ such that
$b = c \rho$ (which always holds if $c = \sigma \sigma^\top$ is
nonsingular). The derivative of the growth rate is $(\nabla \g)_\pi
= b - c \pi$, and it is trivially zero for $\pi \equiv \rho$, which
is the \num portfolio.
\end{ex}

Let us describe in more generality the connection between the \num
and the growth-optimal portfolio, being somewhat informal for the
moment: a vector $\rho \in \C$ maximizes this concave function $\g$
if and only if the directional derivative of $\g$ at the point
$\rho$ in the direction of $\pi - \rho$ is negative for any $\pi \in
\X$. One can use \eqref{eq: growth rate} to compute $(\nabla
\g)_\rho (\pi - \rho)$ and it is straightforward to see that it
turns out to be exactly $\rel(\pi \such \rho)$ of \eqref{eq:
rel_perf}.

Let us try now to be a little more formal. We do not know if we can
differentiate under the integral appearing in equation (\ref{eq:
growth rate}). Even more to the point, we do not know a priori
whether the integral is well-defined: both its positive and negative
parts could be infinite. Non-integrability of the negative part is
not too severe, since one wants to \emph{maximize} $\g$: if a
portfolio $\pi$ results in an integrand whose negative part
integrates to infinity, all vectors $a \pi$ for $a \in [0,1)$ will
lead to a finite result. More problematic is the fact that the
\emph{positive} part can integrate to infinity, especially when one
notices that if this happens for at least one vector $\pi \in \C$,
concavity will imply that it happens for \emph{many} vectors ---
actually for \emph{all} vectors in the relative interior of $\C$,
with the possible exception of those of the form $-a \pi$ for $a
>0$. This problem is related to the one when the expected
log-utility is infinite and one cannot find a unique solution to the
log-utility maximization problem --- see the next subsection
\ref{subsec: log-optimality}.

In the spirit of the above discussion, let us describe a class of
L\'evy measures for which the concave growth rate function
$\g(\cdot)$ of (\ref{eq: growth rate}) \emph{is} well-defined.

\begin{defn} \label{dfn: finite-log-value measure and approximating measures}
A L\'evy measure $\nu$ \textsl{integrates the log}, if $\int \log(1
+ |x|) \indic_{\{ |x| > 1 \}}\nu (\ud x) < \infty$. For any L\'evy
measure $\nu$, a sequence $(\nu_n)_{n \in \Natural}$ of L\'evy
measures that integrate the log with $\nu_n \sim \nu$, whose
densities $f_n := \ud \nu_n / \ud \nu$ satisfy $0 < f_n \leq 1$,
$f_n (x) = 1$ for $|x| \leq 1$, and $\lim_{n \to \infty} \uparrow
f_n = \indic$, will be called an \textsl{approximating sequence}.
\end{defn}

One specific choice for the densities appearing in the definition of
approximating sequence is $f_n(x) = \indic_{\{ |x| \leq 1 \}} +
|x|^{-1/n} \indic_{\{ |x| > 1 \}}$. The sets $\C_0$, $\N$ and $\I$
remain unchanged if we move from the original triplet to any of the
approximating triplets, thus $\I (b, c, \nu) \cap \check{\C} =
\emptyset$ if and only if $\I (b, c, \nu_n) \cap \check{\C} =
\emptyset$ for all $n \in \Natural$.

The problem of the positive infinite value for the integral
appearing in equation (\ref{eq: growth rate}) disappears when the
L\'evy measure $\nu$ integrates the log, and the growth-optimal
portfolio is also the \num portfolio. In the general case, where
$\nu$ might not integrate the log, our strategy will be the
following: solve the optimization problem concerning $\g$ for a
sequence of problems using the approximation described in Definition
\ref{dfn: finite-log-value measure and approximating measures}, and
then show that the corresponding solutions converge to the solution
of the original problem.

\begin{rem}
Even in the unconstrained case \emph{the supermartingale deflator
corresponding to the \num portfolio need not be a martingale, and
can in fact be a strict supermartingale}. Of course, the importance
of supermartingales in utility maximization (after all, we are
basically dealing with log utility here) has been recognized by
Kramkov and Schachermayer \cite{K-S:99}. Hurd \cite{Hurd} gives a
treatment of log-utility in exponential L\'evy models. For
completeness, we give in the next paragraph an elementary example to
illustrate what can go wrong.

Take a one-dimensional L\'evy process with $X$ with $b = 1$, $c=0$
and $\nu (\ud x) = (1+x) \indic_{(-1,1]} (x) \ud x$. One can easily
check that $\C_0 = [-1,1]$ and that $\g'$ (the derivative of $\g$)
is decreasing in $\pi \in (-1,1)$ with $\g' (-1) = + \infty$ and
$\g' (1) = 1/3$. The \num portfolio is $\rho = 1$ and
$(W^\rho)^{-1}$ is a strict L\'evy supermartingale, since $\rel(0
\such 1) = - \g'(1) = - 1/3 < 0$.

The above fact gives some justice to the Esscher transform method in
the proof of Theorem \ref{thm: No Arbitrage for Levy finite time
constrained}, which provides us with a probability measure. The
situation should be contrasted to the continuous-path case of
Example \ref{ex: (NUIP) for continuous} where, in the absence of
constraints, $(W^{\rho})^{-1}$ is a martingale. We also see that we
cannot expect to be able in general to compute the \num portfolio
just by naively trying to solve $\nabla \g (\rho) = \rel(0 \such
\rho) = 0$.
\end{rem}

\subsection{Relative log-optimality and the \num portfolio} \label{subsec: log-optimality}

We rush through the (well-understood) relevance of the \num
portfolio with the \textsl{relatively log-optimal}, i.e., a
portfolio $\rho \in \Pi_\C$ such that $\expec
[\log(W^\pi_T/W^\rho_T)] \leq 0$ (here, it is tacitly assumed that $\expec
\log^+ (W^\pi_T/W^\rho_T) < \infty$), for every $\pi \in \Pi_\C$. A
treatment for the general semimartingale case is given in Karatzas
and Kardaras \cite{KK: num and arbitrage}.

\smallskip

If the \num portfolio $\rho$ exists, then for any other $\pi \in
\Pi_\C$ we have $\expec[W^{\pi}_T / W^{\rho}_T] \leq 1$; applying
Jensen's inequality we get $\expec \log ( W^{\pi}_T/W^{\rho}_T) \leq
0$, i.e., that $\rho$ is relatively log-optimal.

Now, suppose that the \num portfolio does not exist --- according to
Theorem \ref{thm: num iff def-non-empty iff NUPBR iff NUIP}, this
means that we can pick $\xi \in \I \cap \check{\C} \neq \emptyset$.
For any $\rho \in \Pi_\C$, we have $\rho + \xi \in \Pi_\C$ as well;
simple computations, using the fact that $\xi \in \I$, give that the
relative-log-ratio $\log (W_T^{\rho + \xi} / W_T^{\rho})$ is equal
to $( \xi^\top b - \int \xi^\top \hx \nu (\ud x) ) T + \sum_{0 \leq
t \leq T} \log [ 1 + \xi^\top \Delta X_t / (1 + \rho_t^\top \Delta
X_t) ]$, which by Definition \ref{dfn: instant_arb_opport} of
immediate arbitrage opportunities is positive, with positive
probability of being strictly positive; this implies $\expec
\log(W_T^{\rho + \xi} / W_T^{\rho}) > 0$. Thus, if the \num
portfolio does not exist, a relative-log-optimal portfolio cannot
exist either.

\smallskip

The somewhat amazing conclusion from the above discussion above is
that for $\rho \in \Pi_\C$ we have the following equivalence:
\[
\log \Big( \expec \frac{W_T^\pi}{W^\rho_T} \Big) \leq 0, \quad
\textrm{for all } \pi \in \Pi_\C \quad \Longleftrightarrow \quad
\expec \log \Big( \frac{W_T^\pi}{W^\rho_T} \Big) \leq 0, \quad
\textrm{for all } \pi \in \Pi_\C.
\]
Of course, Jensen's inequality gives direction $\Rightarrow$ for any
portfolios $\pi$ and $\rho$ in $\Pi_\C$; the opposite direction
$\Leftarrow$ fails in general for any $\pi$ and $\rho$ in $\Pi_\C$
--- it \emph{will} hold for all $\pi \in \Pi_\C$ if we fix the \emph{specific}
$\rho$ that makes \emph{all} expectations of the relative log-wealth
process non-positive.

\smallskip

If for the relative log-optimal portfolio $\rho$ we have $\expec
\log W^\rho_T < \infty$, then $\rho$ also is the unique log-optimal
portfolio. If $\expec \log W^\rho_T = \infty$, the log-utility
maximization problem has an infinite number of solutions. For an
example where this happens take a one-dimensional L\'evy process
with $b = c = 0$ and a L\'evy measure with density $\nu (\ud x) /
\ud x = \indic_{(-1,1]} (x) + x^{-1} (\log(1+x))^{-2} \indic_{[1,
\infty)} (x)$ --- we have $\C_0 = [0,1]$ and it is easy to check
that $\expec[\log W_T^\pi] = \infty$ for all $\pi \in (0,1)$. For
this example, the problem of maximizing expected log-utility does
not have unique solution. Of course, the \num and relatively
log-optimal portfolios exist and will be unique (and the same).

\section{Finishing the Proof of Theorem \ref{thm: num iff def-non-empty iff NUPBR iff NUIP}} \label{sec: crucial lemma}

The focus of this section is the proof of the following Lemma
\ref{lem: Levy charact of num} which will complete the proof of
Theorem \ref{thm: num iff def-non-empty iff NUPBR iff NUIP}. We
state it separately of everything else because it will also find
good use in Karatzas and Kardaras \cite{KK: num and arbitrage}.

\begin{lem} \label{lem: Levy charact of num}
Let $(b,c,\nu)$ be a L\'evy triplet and $\C$ a closed convex subset
of $\Real^d$. Then, $\I \cap \check{\C} = \emptyset$ \emph{if and
only if} there exists a unique vector $\rho \in \C \cap \N^\bot$
with $\nu [\rho^\top x \leq -1] = 0$ such that $\rel(\pi \such \rho)
\leq 0$ for all $\pi \in \C$.

If $\nu$ integrates the log, the vector $\rho$ above is
characterized as $\rho = \arg \max_{\pi \in \C \cap \N^\bot} \g
(\pi)$. In general, $\rho$ is the limit of solutions to a series of
problems, in which $\nu$ is replaced by a sequence of approximating
measures.
\end{lem}

Although it will come as a result of Theorem \ref{thm: num iff
def-non-empty iff NUPBR iff NUIP}, let us give a quick proof of the
fact that if $\I \cap \check{\C} \neq \emptyset$ then one cannot
find a $\rho \in \C$ such that $\rel( \pi \such \rho) \leq 0$ for
all $\pi \in \C$. To this end, pick a vector $\xi \in \I \cap
\check{\C} \neq \emptyset$, and suppose that $\rho$ satisfied $\rel(
\pi \such \rho) \leq 0$, for all $\pi \in \C$. Since $\xi \in
\check{\C}$, we have $n \xi \in \C$ for all $n \in \Natural$ and the
convex combination $(1-n^{-1}) \rho + \xi \in \C$ too; but $\C$ is
closed, and so $\rho + \xi \in \C$. Easy computations show that
$\rel(\rho + \xi \such \rho)$ is equal to $\xi^\top b - \int
\xi^\top \hx \nu(\ud x) + \int [ \xi^\top x / (1 + \rho^\top x)] \nu
(\ud x)$; this is strictly positive quantity from the definition of
$\xi$. This is a contradiction to $\rho$ satisfying $\rel( \pi \such
\rho) \leq 0$ for all $\pi \in \C$.

\smallskip

We want to prove the converse; namely if $\I \cap \check{\C} =
\emptyset$, then one can find a $\rho$ that satisfies the
requirement of Lemma \ref{lem: Levy charact of num} --- subsections
\ref{subsec: proof_for_bdd_jump_meas} and \ref{subsec: the extension
to non-finite-log-valued Levy measure} are devoted to the proof of
this. In the process we shall need the following simple
characterization of the condition $\I \cap \check{\C} \neq
\emptyset$:

\begin{lem} \label{lem: contradiction to (NUIP)}
If $\C \subseteq \C_0$ and $\xi \in \check{\C} \setminus \N$, then
$\xi \in \I$ if and only if $\rel(0 \such a \xi) \leq 0$ for all $a
\in \Real_+$.
\end{lem}

\proof The fact that $\xi \in \I \cap \check{\C}$ implies $\rel(0
\such a \xi) \leq 0$ for all $a \in \Real_+$ is trivial.

For the converse, let $\xi \in \check{\C} \setminus \N$ satisfy
$\rel(0 \such a \xi) \leq 0$ for all $a \in \Real_+$; we wish to
show that $\xi \in \I$. The second condition of Definition \ref{dfn:
instant_arb_opport} is readily satisfied, since we assume that $\C$
contains the natural constraints. Now, for all $a \in \Real_+$, we
have $-a^{-1} \rel(0 \such a \xi) \geq 0$; writing this down gives
$\xi^\top b - a \xi^\top c \xi + \int [ \xi^\top x / (1 + a \xi^\top
x) - \xi^\top \hx ] \nu (\ud x) \geq 0$. Observe that the integrand
$\xi^\top x / (1 + a \xi^\top x) - \xi^\top \hx$ is $\nu$-integrable
and decreasing in $a$ (remember that $\nu[ \xi^\top x < 0] = 0$), so
we must have $\xi^\top c = 0$ (condition (1) of Definition \ref{dfn:
instant_arb_opport}), which now implies that $\xi^\top b + \int [
\xi^\top x / (1 + a \xi^\top x) - \xi^\top \hx ] \nu (\ud x) \geq
0$. Letting $a \to \infty$ and using the dominated convergence
theorem and  we get condition (3) of Definition \ref{dfn:
instant_arb_opport}, namely $\xi^\top b - \int \xi^\top \hx \nu (\ud
x) \geq 0$. \qed

\medskip

We make one more observation. On several occasions during the course
of the proof we shall use Fatou's lemma in the following form: if we
are given a \emph{finite} measure $\kappa$ and a sequence $(v_n)_{n
\in \Natural}$ of Borel-measurable functions that are
$\kappa$-uniformly bounded from below, then $\int \liminf_{n \to
\infty} v_n(x) \kappa (\ud x) \leq \liminf_{n \to \infty} \int
v_n(x) \kappa (\ud x)$. The finite measures $\kappa$ that we shall
consider will be of the form $\pare{|x| \wedge k}^2 \nu (\ud x)$,
where $k \in \Real_+$ and $\nu$ is our L\'evy measure.

We can now proceed with the proof of the sufficiency of the
condition $\I \cap \check{\C} = \emptyset$ in solving $\rel(\pi
\such \rho) \leq 0$. We shall first do so for the case of a L\'evy
measure that integrates the log,  then extend to the general case.
Throughout the course of the proof we shall be assuming that $\C
\subseteq \C_0$; otherwise, replace $\C$ by $\C \cap \C_0$.

\subsection{Proof of Lemma \ref{lem: Levy charact of num} for a
L\'evy measure that integrates the log}\label{subsec: proof_for_bdd_jump_meas}

We are trying to show (1) $\Rightarrow$ (2) of Lemma \ref{lem: Levy
charact of num}, so let us assume $\I \cap \check{\C} = \emptyset$.
For this subsection we also make the assumption $\int_{\{ |x| > 1
\}} \log(1 + |x|) \nu (\ud x) < \infty$.

Recall from subsection \ref{subsec: growth-optimal portfolios} the
growth rate function $\g$ of (\ref{eq: growth rate}). This is a
concave function on $\C$, it is well-defined, in the sense that we
always have $\g(\pi) < +\infty$ for $\pi \in \C$ and upper
semi-continuous on $\C$ (the last two facts follow because $\nu$
integrates the log). Of course, $\g$ can take the value $- \infty$
on the boundary of $\C$.

Set $\g_* := \sup_{\pi \in \C} \g (\pi)$, and let $(\rho_n)_{n
\in \Natural}$ be a sequence of vectors in $\C$ with $\lim_{n \to
\infty} \g(\rho_n) = \g_*$. Since for any $\pi \in \C$ and any
$\zeta \in \N$ we have $\g(\pi + \zeta) = \g(\pi)$, we can choose
the sequence $\rho_n$ to take values on the subspace $\N^\bot$ (it
would be useful to recall the discussion of Remark \ref{rem: does
not hurt to consider non-degeneracy}).

\smallskip

We first want to show that the sequence $(\rho_n)_{n \in \Natural}$
of vectors of $\C \cap \N^\bot$ is bounded; then we shall be able to
pick a convergent subsequence. Suppose then on the contrary that
$(\rho_n)_{n \in \Natural}$ unbounded, and without loss of
generality suppose also that the sequence of unit-length vectors
$\xi_n := \rho_n / |\rho_n|$ converges to a unit-lenth vector $\xi
\in \N^\bot$ (picking a subsequence otherwise). We shall use Lemma
\ref{lem: contradiction to (NUIP)} applied to the vector $\xi$ and
show that $\xi \in \I \cap \check{\C}$, contradicting condition (1)
of Lemma \ref{lem: Levy charact of num}.

Start by picking any $a \in \Real_+$; for all large enough $n \in
\Natural$ we have $a \xi_n \in \C$, and since $\C$ is closed we have
$a \xi \in \C$ as well, which implies $\xi \in \check{\C}$ (since $a
\in \Real_+$ is arbitrary). We have $\xi \in \check{\C} \setminus
\N$, and only need to show $\rel(0 \such a \xi) \leq 0$. For this,
we can assume that the sequence $(\rho_n)_{n \in \Natural}$ is
picked in such a way that the functions $[0,1] \ni u \mapsto \g(u
\rho_n)$ are increasing; otherwise, replace $\rho_n$ by the vector
$u \rho_n$ for the choice of $u \in [0,1]$ that maximizes $[0,1] \ni
u \mapsto \g(u \rho_n)$. This would imply that eventually, for all
large enough $n \in \Natural$ we have $\rel(0 \such a \xi_n) \leq
0$; this means
\[
\int \bra{ \frac{- \xi_n^\top x}{1 + a \xi_n^\top x} + \xi_n^\top \hx } \nu (\ud x) \leq
\xi_n^\top b - a \xi_n^\top c \xi_n.
\]
If we can show that we can apply Fatou's lemma to the quantity on
the left-hand-side of this inequality, we get the same inequality
with $\xi$ in place of $\xi_n$ and so $\rel(0 \such a \xi) \leq 0$;
an application of Lemma \ref{lem: contradiction to (NUIP)} shows
that $\xi \in \I \cap \check{\C}$, contradicting condition (1) of
Lemma \ref{lem: Levy charact of num}.

To show that we can apply Fatou's lemma, let us show that the
integrand is bounded from below for the finite measure $\pare{ |x|
\wedge k }^2 \nu(\ud x)$ with $k := 1 \wedge (2a)^{-1}$. Since
$\xi_n^\top x/(1 + a \xi_n^\top x) \leq a^{-1}$ and $\abs{\xi_n^\top
x} \leq |x|$, the integrand is uniformly bounded from below by
$-(a^{-1} + 1)$, and we only need consider what happens on the set
$\{ |x| \leq k \}$; there, the integrand is equal to $- a
(\xi_n^\top x)^2/(1 + a \xi_n^\top x)$, which cannot be less than
$-2 a |x|^2$ and we are done.

\smallskip

We now know that $(\rho_n)_{n \in \Natural}$ is bounded in
$\Real^d$; without loss of generality, suppose that $(\rho_n)_{n \in
\Natural}$ converges to a point $\rho \in \C$ (otherwise, choose a
convergent subsequence). The concavity of $\g$ implies that $\g_*$
is a finite number and it is obvious from continuity that $\g(\rho)
= \g_*$. Of course, we have that $\nu \bra{\rho^\top x \leq -1} =
0$, otherwise $\g(\rho) = -\infty$.

Pick now any $\pi \in \C^\diamond := \{ \pi \in \C \such \nu[
\pi^\top x \leq -u ] = 0 \text{ for some } u < 1 \}$, then it is
clear that $g(\pi) > -\infty$. If follows that the mapping $[0,1]
\ni u \mapsto \g( \rho + u (\pi - \rho) )$ is well-defined (i.e.,
real-valued), concave and decreasing, so that the right-derivative
at $u=0$ should be negative; this derivative is just $\rel(\pi \such
\rho)$, so we have $\rel(\pi \such \rho) \leq 0$ for $\pi \in
\C^\diamond$.

The extension of the inequality $\rel(\pi \such \rho) \leq 0$ for
all $\pi \in \C$ now follows easily. Indeed, if $\pi \in \C$, then
for $0\leq u < 1$ we have $u \pi \in \C^\diamond$ and $\rel(u \pi
\such \rho) \leq 0$; by using Fatou's lemma one can easily check
that we also have $\rel(\pi \such \rho) \leq 0$. \qed

\subsection{The extension to general L\'evy measures} \label{subsec: the extension to non-finite-log-valued Levy measure}

We now have to extend the result of the previous subsection to the
case where $\nu$ does not necessarily integrate the log. Recall from
Definition \ref{dfn: finite-log-value measure and approximating
measures} the use of the approximating triplets $(b,c,\nu_n)$, where
for every $n \in \Natural$ we define the measure $\nu_n (\ud x) :=
f_n (x) \nu(\ud x)$; all these measures integrate the log. We assume
throughout that $\I \cap \check{\C} = \emptyset$.

We remarked that the sets $\N$ and $\I$ remain invariant if we
change the L\'evy measure from $\nu$ to $\nu_n$. Then, since we have
$\I(b,c,\nu_n) \cap \check{\C} = \emptyset$, the discussion in the
previous section, gives us unique vectors $\rho_n \in \C \cap
\N^\bot$ such that $\rel_n (\pi \such \rho_n) \leq 0$ for all $\pi
\in \C$, where $\rel_n$ is associated with the triplet $(b, c,
\nu_n)$.

\smallskip

As before, the constructed sequence $(\rho_n)_{n \in \Natural}$ is
bounded. To prove it, we shall use Lemma \ref{lem: contradiction to
(NUIP)} again, in the exact same way that we did for the case of a
measure that integrates the log. Assume by way of contradiction that
$(\rho_n)_{n \in \Natural}$ is not bounded. By picking a subsequence
if necessary, assume without loss of generality that $|\rho_n|$
diverges to infinity. Now, call $\xi_n := \rho_n / |\rho_n|$. Again,
by picking a further subsequence if the need arises, assume that
$\lim_{n \to \infty} \xi_n = \xi$, where $\xi$ is a unit vector in
$\N^\bot$. Since $\rho_n \in \C$ for all $n \in \Natural$ it follows
that $a \xi \in \C$ for all $a \in \Real_+$, i.e., $\xi \in
\check{\C} \setminus \N$. We know that for sufficiently large $n \in
\Natural$, we have that $\rel_n (0 \such a \xi_n) \leq 0$;
equivalently $\int [- \xi_n^\top x f_n(x)/(1 + a \xi_n^\top x)  +
\xi_n^\top \hx ] \nu (\ud x) \leq \xi_n^\top b - a \xi_n^\top c
\xi_n$. The situation is exactly the same as in the proof in the
case of a measure that integrates the log, but for the appearance of
the density $f_n(x)$ which can only have a positive effect on any
lower bounds that we have established there, since $0 < f_n \leq 1$.
We show that the integrand is bounded from below for the finite
measure $\pare{ |x| \wedge k }^2 \nu(\ud x)$ with $k = 1 \wedge
(2a)^{-1}$, thus we can apply Fatou's lemma to the left-hand-side of
this inequality to get the same inequality with $\xi$ in place of
$\xi_n$, and so $\rel(0 \such a \xi) \leq 0$. Invoking Lemma
\ref{lem: contradiction to (NUIP)}, we arrive at a contradiction
with the assumption $\I \cap \check{\C} = \emptyset$.

\smallskip

Now that we know that $(\rho_n)_{n \in \Natural}$ is a bounded
sequence, we can assume that it converges to a point $\rho \in \C
\cap \N^\bot$, picking a subsequence if needed. We shall show that
$\rho$ satisfies $\rel(\pi \such \rho) \leq 0$ for all $\pi \in \C$.
Pick any $\pi \in \C$; we know that we have
\[
\int \bra{ \frac{(\pi - \rho_n)^\top x}{1 + \rho_n^\top x} f_n(x) - (\pi - \rho_n)^\top \hx } \nu
(\ud x) \leq - (\pi - \rho_n)^\top b + (\pi - \rho_n)^\top c \rho_n
\]
for all $n \in \Natural$. Yet once more, we shall use Fatou's lemma
on the left-hand-side to get to the limit the same inequality with
$\rho_n$ and $f_n(x)$ being replaced by $\rho$ and $1$ respectively;
in other words, we get $\rel(\pi \such \rho) \leq 0$ for all $\pi
\in \C$.

To justify the use of Fatou's lemma, we shall show that the
integrands are uniformly bounded from below for the finite measure
$(|x| \wedge k)^2 \nu(\ud x)$, where $k := 1 \wedge (2 \sup_{n \in
\Natural} |\rho_n|)^{-1}$ is a strictly positive number from the
boundedness of $(\rho_n)_{n \in \Natural}$. First, observe that the
integrands are uniformly bounded by $-1 - \sup_{n \in \Natural} |\pi
- \rho_n|$, which is a finite number. Thus, we only need worry about
the set $\{ |x| \leq k \}$. There, the integrands are equal to $(\pi
- \rho_n)^\top x (\rho_n^\top x)/(1 + \rho_n^\top x)$; this cannot
be less than $-2 \sup_{n \in \Natural} (|\pi - \rho_n| |\rho_n|)
|x|^2$, and Fatou's lemma can be used.

\smallskip

Up to now we have shown that $\rel(\pi \such \rho) \leq 0$ for all
$\pi \in \C$ for the limit $\rho$ of a subsequence of $(\rho_n)_{n
\in \Natural}$. Nevertheless, carrying the previous steps we see
that \emph{every} subsequence of $(\rho_n)_{n \in \Natural}$ has a
further convergent subsequence whose limit $\hat{\rho} \in \C \cap
\N^\bot$ satisfies $\rel(\pi \such \hat{\rho}) \leq 0$ for all $\pi
\in \C$. The uniqueness of $\rho \in \C \cap \N^\bot$ that satisfies
$\rel(\pi \such \rho) \leq 0$ for all $\pi \in \C$ gives that
$\hat{\rho} = \rho$, and we conclude that the whole sequence
$(\rho_n)_{n \in \Natural}$ converges to $\rho$. \qed

\appendix

\section{Facts Regarding L\'evy Processes} \label{sec: appendix on levy}

We hereby collect some results that are used within the text; they are mostly simple consequences
of the definition of a L\'evy process; we include them for completeness, since they might not be
part of the usual treatment in textbooks.

\medskip

First of all, L\'evy process have the following property, which already points out in some way the
fact that ``if there is arbitrage it should be an \emph{increasing} profit'':

\begin{lem} \label{lem: pos is incr for Levy}
If for some one-dimensional L\'evy process $L$ and some time $T > 0$ we have $L_T \geq 0$,
$\prob$-a.s., then $L$ is actually an increasing process.
\end{lem}

\proof Write $L_T = L_{T/2} + L'_{T/2}$, where $L'_{T/2}$ is
independent of, and has the same distribution as $L_{T/2}$. Then, $0
= \prob[L_T < 0] = \prob[ L_{T/2} < - L'_{T/2}] \geq \prob[L_{T/2} <
0, L'_{T/2} < 0] = (\prob[L_{T/2} < 0])^2$, hence $\prob[L_{T/2} <
0] = 0$. Continuing like this and using the stationary-increments
property of $L$ we get $\prob [L_t < 0] = 0$ for all $t \in
\mathbb{D} := \{k T / 2^n \such n \in \Natural, \ k = 0, \ldots, 2^n
\}$. The stationarity of increments of $L$ couple with the
countability of $\mathbb{D}$ implies that $\prob[L_s \leq L_t
\textrm{ for all } s \in \mathbb{D}, \ t \in \mathbb{D} \textrm{
with } s < t] = 1$; then, right-continuity of $L$ will give us that
the latter is an increasing process. \qed

\medskip

A $\mathbf{F}$-L\'evy process $X$ is \emph{regenerating} at every
stopping time $\sigma$ --- this means that on $\{ \sigma < \infty
\}$ the process $Y := (X_{\sigma + s} - X_\sigma)_{s \in \Real_+}$
is an $\mathbf{G}$-L\'evy process, independent of $\F_{\sigma}$,
where we set $\mathcal{G}_s := \F_{\sigma + s}$ for all $s \in
\Real_+$. If $\tau$ is an $\mathbf{F}$-stopping time with $\sigma
\leq \tau$, $\prob$-a.s., then the random time $\tau - \sigma$ is an
$\mathbf{G}$-stopping time and we obviously have $\Delta Y_{\tau -
\sigma} = \Delta X_\tau \indic_{\{ \sigma < \tau \}}$. These remarks
will be used in the proof of the result below which states that the
jump-size at a stopping time is independent of whatever has happened
\emph{strictly} before that stopping time. This ``strict history''
notion is formalized by introducing the $\sigma$-algebra $\F_{\tau -
}$ of events strictly prior to $\tau$, that is the smallest
$\sigma$-algebra generated by the class $\A_{\tau-} := \F_0 \cup \{
B \cap \{ t < \tau\} \such B \in \F_t \textrm{ for some } t \in
\Real_+ \}$.

\begin{lem} \label{lem: jump indep of strict hist}
If $X$ is an $\mathbf{F}$-L\'evy process for some filtration
$\filtration$, then for any stopping time $\tau$, the jump $\Delta
X_\tau \indic_{ \{ \tau < \infty \}}$ is independent of $\F_{\tau
-}$.
\end{lem}

\proof The class $\A_{\tau-}$ defined above is closed under
intersection and generates $\F_{\tau -}$. Therefore, it suffices to
prove that all $A \in \A_{\tau - }$ are independent of $\Delta
X_{\tau}$. For $A \in \F_0$ this is trivial. Thus, consider $A = B
\cap \{ t < \tau\}$ for some $B \in \F_t$. Let $\sigma := \tau
\wedge t$; we have $\sigma \leq \tau$ and the regenerating property
of L\'evy processes implies that $Y := (X_{\sigma + s} -
X_\sigma)_{s \in \Real_+}$ is an $\mathbf{G}$-L\'evy process,
independent of $\F_{\sigma}$, where again $\mathcal{G}$ was defined
above. These considerations give us that
\[
\prob[A \cap \{ \Delta X_{\tau} \in D \}] = \prob[B \cap \{ t <
\tau\} \cap \{ \Delta Y_{\tau - \sigma} \in D \}] = \prob[B \cap \{
t < \tau\}] \prob [Y_{\tau - \sigma} \in D]
\]
for all $D \in \B(\Real^d)$; the last term above is just $\prob[A]
\prob[ \Delta X_{\tau} \in D]$, and the claim follows. \qed

\medskip

If the L\'evy measure $\nu$ of the L\'evy process $X$ has finite
mass ($\nu(\Real^d) < \infty$), then one can represent $X$ in the
following form: $X_t = \widetilde{b} t + \sigma \beta_t +
\sum_{i=1}^{N_t} Y_i$, where $N$ is a Poisson process with rate
$\nu(\Real^d)$ and $Y_i$ is a sequence of independent and
identically distributed random variables with distribution
$\nu(\cdot) / \nu(\Real^d)$, further independent of $N$. In that
case we can define the time of the $n^{\textrm{th}}$ jump of $X$ via
$\tau_n := \inf \{t \in \Real_+ \such N_t = n\}$. The independence
of $N$ and $(Y_n)_{n \in \Natural}$ gives that $\Delta X_{\tau_n}$
has the distribution of $Y_1$ and is independent of $\tau_n$. For
general stopping times $\tau$ with $\prob[\Delta X_\tau \neq 0] = 1$
we cannot of course expect that $\Delta X_{\tau}$ has the same
distribution as $Y_1$, since we might be sampling the paths in a
biased way; for example if $D$ is a Borel subset of $\Real^d
\setminus \{0\}$ and $\tau := \inf \{ t \in \Real_+ \such \Delta X_t
\in D \}$ then $\Delta X_\tau$ is only supported on $D$.
Nevertheless, if the decision on whether to stop at the
$n^{\textrm{th}}$ jump of $X$ or not is depending \emph{only} on
information collected \emph{strictly before} $\tau_n$, the fact that
$\Delta X_\tau$ has the same distribution as $Y_1$ is still valid.

\begin{lem} \label{lem: sampled jump has same distr as first}
If the L\'evy measure $\nu$ of the L\'evy process $X$ is such that
$\nu(\Real^d) < \infty$, and with the notation set above, consider
the stopping time $\tau := \bigwedge_{n =1}^{\infty}
(\tau_n)_{A_n}$, where we have set as usual $(\sigma)_A := \sigma
\indic_A + \infty \indic_{\Omega \setminus A}$ for a random time
$\sigma$ and $A \subseteq \Omega$. If $A_n \in \F_{\tau_n -}$ for
all $n \in \Natural$, then, conditional on $\{ \tau < \infty \}$,
$\Delta X_{\tau}$ is identically distributed as $Y_1$.
\end{lem}

\proof Observe first of all that we can assume that the sequence
$(A_n)_{n \in \Natural}$ consists of disjoint sets; otherwise, we
can replace $A_n$ by $A_n \setminus (\bigcup_{i < n} A_i)$; these
sets are still in $\F_{\tau_n -}$, they are disjoint and $\tau$ is
still given by the same formula $\tau = \bigwedge_{n =1}^{\infty}
(\tau_n)_{A_n}$. We obviously have $\{ \tau < \infty \} = \bigcup_{n
\in \Natural} A_n$. Pick any Borel-measurable $g : \Real^d \mapsto
\Real_+$; writing $g(\Delta X_\tau) = \sum_{n=1}^\infty g(\Delta
X_{\tau_n}) \indic_{A_n}$ and observing that the previous Lemma
\ref{lem: jump indep of strict hist} implies $\expec[g(\Delta
X_{\tau_n}) \indic_{A_n}] = \expec[g(\Delta X_{\tau_n})] \prob[A_n]$
for all $n \in \Natural$, we get
\[
\expec[g(\Delta X_\tau) \indic_{\{\tau < \infty\}} ] =
\sum_{n=1}^\infty \expec [g(\Delta X_{\tau_n})] \prob [ A_n ] =
\sum_{n=1}^\infty \expec [g(Y_1)] \prob [ A_n ] = \expec [g(Y_1)]
\prob [ \tau < \infty ];
\]
in other words, $\expec[g(\Delta X_\tau) \such \tau < \infty ] =
\expec [g(Y_1)]$, i.e., $\Delta X_\tau$ is identically distributed
as $Y_1$. \qed



\begin{thebibliography}{9}

\bibitem{Algoet-Cover} \textsc{P. Algoet, Tom M. Cover} (1988). ``Asymptotic optimality
and asymptotic equipartition property of log-optimal investment'',
Annals of Probability {\bf 16}, pp. 876--898.

\bibitem{Ansel-Stricker} \textsc{Jean-Pascal Ansel,
Christophe Stricker} (1994). ``Couverture des actifs contigents et
prix maximum'', Annales de l' Institute Henri Poincar\'e {\bf 30},
p. 303--315.

\bibitem{Becherer} \textsc{Dirk Becherer} (2001). ``The num\'eraire portfolio for
unbounded semimartingales'', Finance and Stochastics 5, p. 327--341.

\bibitem{CGMY} \textsc{ Peter Carr, H\'elyette Geman, Dilip B. Madan, Marc Yor} (2002).
``The Fine Structure of Asset Returns: An Empirical Investigation'',
Journal of Business 75, p. 305–-332.

\bibitem{Cherny : NGA} \textsc{Alexander S. Cherny} (2005). ``General arbitrage pricing model: probability
approach'', to be published in Lecture Notes in Mathematics

\bibitem{Cherny : Levy process} \textsc{Alexander S. Cherny, Albert N. Shiryaev}
(2002). ``Change of time and measure for L\'evy processes''.
Lectures at the Summer School ``From Levy processes to
semimartingales: recent theoretical developments and applications in
finance'' (Aarhus).


\bibitem{Cont-Tankov: Levy} \textsc{Rama Cont, Peter Tankov} (2004). ``Financial Modelling With Jump Processes'',
Chapman \& Hall/CRC.

\bibitem{DS: FTAP locally bdd} \textsc{Freddy Delbaen, Walter Schachermayer} (1994). ``A
General Version of the Fundamental Theorem of Asset Pricing'',
Mathematische Annalen 300, p. 463--520.

%
%
\bibitem{DS: Bessel} \textsc{Freddy Delbaen, Walter Schachermayer} (1995).
``Arbitrage Possibilities in Bessel Processes and their Relations to
Local Martingales'', Probability Theory and Related Fields 102,
n$^\text{o}$ 3, pp. 357--366.

\bibitem{DS: FTAP unbounded} \textsc{Freddy Delbaen, Walter Schachermayer} (1998). ``The Fundamental Theorem of Asset
Pricing for Unbounded Stochastic Processes'', Mathematische Annalen
312, n$^\text{o}$ 2, p. 215--260.


\bibitem{Jacod - Shiryaev} \textsc{Jean Jacod, Albert N. Shiryaev} (2003). ``Limit Theorems for Stochastic
Processes'', Second Edition. Springer.

%

\bibitem{Ebe-Jac: range of option prices} \textsc{Ernst Eberlein, Jean Jacod} (1997).
``On the Range of Options Prices'', Finance and Stochastics, vol. 1,
pp. 131--140.

\bibitem{Ebe-Ke-Pra} \textsc{Ernst Eberlein, Ulrich Keller, Karsten
Prause} (1998). ``New Insights into Smile, Mispricing and Value at
Risk: The Hyperbolic Model'', The Journal of Business, vol. 71,
n$^\text{o}$ 3, pp. 371--406.

\bibitem{Esc-Schw} \textsc{Felix Esche, Martin Schweizer} (2005). ``Minimal entropy preserves the L\'evy property: how
and why'', Stochastic Processes and their Applications 115, pp.
299–-327.

\bibitem{Foldes} \textsc{Lucien P. Foldes} (1991). ``Optimal Sure Portfolio Plans'',
Mathematical Finance 1, pp. 15–-55.

\bibitem{Fuj-Miya} \textsc{Tsukasa Fujiwara, Yoshio Miyahara} (2003) ``The minimal entropy martingale measures
for geometric L\'evy processes'', Finance and Stochastics 7, pp.
509–-531.

\bibitem{Goll - Kallsen: log-optimal} \textsc{T. Goll, J.
Kallsen} (2003).``A Complete Explicit Solution to the Log-Optimal
Portfolio Problem'', The Annals of Applied Probability {\bf 13}, p.
774--799.

\bibitem{Hub-Sga} \textsc{Friedrich Hubalek, Carlo Sgarra} (2006) ``Esscher transforms and the minimal entropy martingale measure for
exponential L\'evy models'', Quantitative Finance, Volume 6, Issue
2, pp. 125--145.

\bibitem{Hurd} \textsc{T. R. Hurd} (2004) ``A note on log-optimal portfolios in exponential L\'evy markets'',
Statistics and Decisions, Volume 22, Issue 3, pp. 225--233.



\bibitem{Jacub: option pricing} \textsc{Paulius Jacub\.enas} (2002).
``On Option Pricing in Certain Incomplete Markets'', Proceedings of
the Steklov Institute of Mathematics, vol. 237, pp. 114--133.


\bibitem{KK: num and arbitrage} \textsc{Ioannis Karatzas, Constantinos Kardaras} (2006).
``The Num\'eraire Portfolio in Semimartingale Financial Models'', to
appear in ``Finance and Stochastics''.

\bibitem{K: thesis} \textsc{Constantinos Kardaras} (2006).
``The \num portfolio and arbitrage in semimartingale models of
financial markets''. Ph.D. Dissertation, Columbia University

\bibitem{K-S:99} \textsc{Dmitry Kramkov, Walter Schachermayer} (1999). ``The
Asymptotic Elasticity of Utility functions and Optimal Investment in
Incomplete Markets'', The Annals of Applied Probability, Vol 9,
n$^\text{o}$ 9, p. 904--950.


\bibitem{Rogers} \textsc{L. C. G. Rogers} (1994). ``Equivalent Martingale
measures and no Arbitrage'', Stochastics and Stochastics Reports 51,
n$^\text{os}$ 1--2, pp. 41--50.

\bibitem{Sato} \textsc{Ken-Iti Sato} (1999).
``L\'evy Processes and Infinitely Divisible Distributions'',
Cambridge University Press.

\bibitem{Selivanov} \textsc{A. V. Selivanov} (2005).
``On the Martingale Measures in Exponential L\'evy Models'', Theory
of Probability and its Applications, volume 49, issue 2, pp.
261--274.

%

\bibitem{Yan} \textsc{Jia An Yan} (1998).
``A new look at the fundamental theorem of asset pricing'', Journal
of the Korean Mathematical Society, volume 35, n$^\text{o}$ 3, pp.
659--673.

\end{thebibliography}
\end{document}